\documentclass[aip, reprint, amsmath,amssymb,citeautoscript]{revtex4-1}
\pdfoutput=1
\usepackage{graphicx}
\usepackage{dsfont}
\usepackage{xcolor}

\usepackage{float}

\begin{document}

%\preprint{}

\title{Breaking Symmetries of the Reservoir Equations in Echo State Networks} %Title of paper



%\author{}
%\thanks{}
%\affiliation{}

\author{Joschka Herteux}
\email{joschka.herteux@dlr.de}
\affiliation{Institut f{\"u}r Materialphysik im Weltraum, Deutsches Zentrum f{\"u}r Luft- und Raumfahrt,
M{\"u}nchner Str. 20, 82234 Wessling, Germany}
\author{Christoph R{\"a}th}
\email{christoph.raeth@dlr.de}
\affiliation{Institut f{\"u}r Materialphysik im Weltraum, Deutsches Zentrum f{\"u}r Luft- und Raumfahrt,
M{\"u}nchner Str. 20, 82234 Wessling, Germany}


\date{\today}

\begin{abstract}
Reservoir computing has repeatedly been shown to be extremely successful in the prediction of nonlinear 
time-series. However, there is no complete understanding of the proper design of a reservoir yet. We find 
that the simplest popular setup has a harmful symmetry, which leads to the prediction of what we 
call \textit{mirror-attractor}. We prove this analytically. Similar problems can arise in a 
general context, and we use them to explain the success or failure of some designs. The symmetry
is a direct consequence of the hyperbolic tangent activation function. Further, four ways to 
break the symmetry are compared numerically: A bias in the output, a shift in the input, a quadratic term 
in the readout, and a mixture of even and odd activation functions. Firstly, we test their 
susceptibility to the mirror-attractor. Secondly, we evaluate their performance on the task of 
predicting Lorenz data with the mean shifted to zero. The short-time prediction is measured with
the forecast horizon while the largest Lyapunov exponent and the correlation dimension are used 
to represent the climate. Finally, the same analysis is repeated on a combined dataset of the 
Lorenz attractor and the Halvorsen attractor, which we designed to reveal potential problems 
with symmetry. We find that all methods except the output bias are able to fully break the 
symmetry with input shift and quadratic readout performing the best overall.
\end{abstract}

%\pacs{XXX}% insert suggested PACS numbers in braces on next line

\maketitle %\maketitle must follow title, authors, abstract and \pacs

\begin{quotation}
Reservoir computing describes a kind of recurrent neural network, which has been very successful in the
prediction of chaotic systems. However, the details of its inner workings have yet to be fully understood.
One important aspect of any neural network is the 
activation function. Even though its effects have been extensively studied in other Machine Learning
techniques, there are still open questions in the context of reservoir computing. Our research aims to
fill this gap. We prove
analytically that an antisymmetric activation function like the hyperbolic tangent leads to 
a disastrous symmetry in a popular setup we call simple ESN. This leads the reservoir to learn an
inverted version of the training data we call \textit{mirror-attractor}, which we demonstrate numerically.
This heavily perturbs any 
prediction, especially if the mirror-attractor overlaps with the real attractor. Further, we 
compare four different ways to break the symmetry. We test 
numerically if they tend to learn the mirror-attractor and test their performance
on two tasks where the simple ESN fails. We find that three of them are able 
to fully break the symmetry.

\end{quotation} 

\section{Introduction}

Machine Learning (ML) has shown to be tremendously successful in categorization and recognition tasks
and the use of ML algorithms has become common in technical devices of daily living. But the application of ML also pervades more and more areas 
of science including research on complex systems.  For a very recent collection see \cite{tang2020} and references therein.\\
In nonlinear dynamics ML-based methods have recently attracted a lot of attention, because it was demonstrated 
that the exact short term prediction of nonlinear system can be significantly improved. Furthermore, it was shown that 
ML techniques also allow for a very accurate reproduction of the long term properties ("the climate") 
of complex systems \cite{lu2018attractor, pathak2017using}. Several ML methods like deep feed-forward artificial neural network (ANN), 
recurrent neural network (RNN) with long short-term memory (LSTM) 
and reservoir computing (RC) fulfill the prediction tasks \cite{vlachas2019backpropagation,chattopadhyay2019data}. 
RC has attracted most attention. It is a machine learning method that has been independently 
discovered as Liquid
State Machines (LSM) by Maass \cite{maass02} and as echo state networks (ESN) by Jaeger \cite{jaeger2001echo}. We focus here on the ESN
approach, which falls under the category of Recurrent Neural Networks (RNN). The main difference 
to other RNNs such as LSTMs is that in RC only the last layer is explicitly 
trained via linear regression.
Instead of hidden layers it uses a  so-called \textit{reservoir}, which in the case of the ESN is 
typically 
a network with recurrent connections.\\
The popularity of RC has several reasons. First, RC often shows superior performance. 
Second, ESNs offer conceptual advantages. As only the output layer is explicitly trained, 
the number of weights to be adjusted is very small. Thus, the training of ESNs is comparably 
transparent, extremely CPU-efficient (orders of magnitude faster than for ANNs) and the 
vanishing-gradient-problem is circumvented. Furthermore, small, smart and energy-efficient 
hardware implementations using  photonic systems\cite{van2017advances}, spintronic systems 
\cite{prychynenko2018magnetic} and many more are 
conceivable and being developed \cite{tanaka2019recent} (and references therein).\\
Ongoing research is focused on identifying the  necessary conditions for a good 
reservoir. Recent studies focused mainly on the influence of the size and topology of the reservoir
on the prediction capabilities \cite{griffith2019forecasting, carroll2019network, haluszczynski2019good,
haluszczynski2020reducing, carroll2020path}.
Less attention has so far been paid on the role of activation and onto the overall performance of 
RC.\\ 
In this paper we study in detail the sensitivity of RC to symmetries in the activation function. 
We reveal that previously reported shortcomings for simple ESNs can unambiguously be attributed 
to  symmetry properties of the activation function (and not of the input signal). We propose and 
assess four different methods to break the symmetries that were developed for obtaining more 
reliable prediction results. \\
The paper is organized as follows: Sec. \ref{sec:Methods}
first discusses the different measures used and the two test systems: the Lorenz and 
the Halvorsen equations. Afterwards the different ESN designs used in this study are introduced 
and the symmetry of the simple ESN is proven. In Sec. \ref{sec:Results} the three numerical
experiments we conducted and their results are presented. Finally we discuss our findings in Sec. 
\ref{sec:Conclusion}.

%%%%%%%%%%%%%%%%%%%%%%%%%%%%%%%
\section{Methods}
\label{sec:Methods}

\subsection{Measures and System Characteristics}
\label{sec:Measures}

\subsubsection{Forecast Horizon} 

As in \cite{haluszczynski2019good, haluszczynski2020reducing} we use the \textit{forecast horizon}
to measure the quality of short-time 
predictions. It is defined as the time between the start of a prediction and the point where it 
deviates from the test data more than a fixed threshold. The exact condition reads
\begin{eqnarray}
| \textbf{v}(t) - \textbf{v}_{R}(t) | > \boldsymbol{\delta} \ ,
\label{eq:fchor}
\end{eqnarray} 
where the norm is taken elementwise.
Due to the chaotic nature of our training
data, any small perturbation will usually grow exponentially with time. Thus, this indicates the end of a 
reliable prediction of the actual trajectory. The measure is not very sensitive to the exact value
of the threshold for the same reason.\\
The threshold generally depends on the direction and we use $\boldsymbol{\delta} = (5.8, 8.0, 6.9)^{T}$ for the Lorenz system without
preprocessing. In general the values of $\delta$ are chosen to be approximately 15\% of the spatial 
extent of the respective attractor in the given direction. This is useful if the dynamics of a system 
takes place on different lengthscales.

\subsubsection{Correlation Dimension} 

To evaluate the climate of a prediction we use two measures. To understand the 
structural complexity of the attractor it is interesting to look at the correlation dimension.
This is a way to quantify the dimensionality of the space populated by the trajectory \cite{grassberger1983measuring}.
The correlation dimension is based on the discrete form of the correlation integral 
\begin{eqnarray}
\begin{aligned}
C(r) &= \lim\limits_{N \rightarrow \infty}{\frac{1}{N^2}\sum^{N}_{i,j=1}\theta(r- | \textbf{x}_{i} - \textbf{x}_{j}  |)} \ ,
\label{eq:corrintegral}
\end{aligned}
\end{eqnarray} 
which returns the fraction of pairs of points that are closer than the threshold distance $r$.
$\theta$ represents the Heaviside function. 
The correlation dimension is then defined by the relation
\begin{eqnarray}
C(r) \propto r^{\nu} 
\label{eq:corrdim}
\end{eqnarray} 
as the scaling exponent $\nu$. For a self-similar strange attractor this relation holds in
some range of $r$, which needs to be properly calibrated. Here we adjusted it for every given 
problem beforehand on simulated data, which is not used for training or testing. We note that
precision is 
not essential here, since we are only interested in comparisons and not in absolute values.\\
To get the correlation dimension for a given dataset we use the Grassberger Procaccia algorithm
\cite{grassberger83a}. 

\subsubsection{Largest Lyapunov Exponent} 

The second measure we use to evaluate the climate is the largest Lyapunov exponent. In contrast 
to the correlation dimension it is indicative of the development of the system in time. A 
$d$-dimensional chaotic system is characterized by $d$ Lyapunov exponents of which at least one 
is positive. They describe the average rate of exponential growth of a small perturbation in each
direction in phase space. The largest Lyapunov exponent $\lambda$ is the one associated with the 
direction of the fastest divergence. 
\begin{eqnarray}
d(t) =  C e^{\lambda t} \ .
\label{eq:lyapunov}
\end{eqnarray} 
Since it dominates the dynamics it has a special significance. It can be calculated from data 
with relative ease by using the Rosenstein algorithm \cite{rosenstein1993practical}.
It is also possible to determine the complete Lyapunov spectrum from the equations, which we
have access to for our testdata as well as for our ESNs \cite{pathak2017using, sandri1996numerical}. 
However, we found that the comparison is clearer in our case with the data-driven approach, because it 
is completely independent of details of the system, e.g. the question if it is discrete or continuous.
This method is also computationally less costly. \\
We can further define the \textit{Lyapunov time} $\tau_{\lambda} = \frac{1}{\lambda}$ as characteristic 
timescale of a system. We use $\tau_{\lambda} =\frac{1}{0.87} \approx 1.15$ for the Lorenz system and 
$\tau_{\lambda} =\frac{1}{0.74} \approx 1.35$ for the Halvorsen system, based on our measurements in table
\ref{tab:halorenz}.
%%%%%%

\subsection{Lorenz and Halvorsen system}
\label{sec:Lorenz}

%%%%%%
A standard example of a chaotic attractor is provided by the Lorenz system\cite{lorenz1963deterministic}. It is 
widely used as a test case for prediction of such systems with RC. It is defined 
by the equations
\begin{eqnarray}
\begin{aligned}
\dot x &= \sigma (y-x) \\
\dot y &= x (\rho-z)-y \\
\dot z &= x y - \beta z + x \ ,
\label{eq:lorenz}
\end{aligned}
\end{eqnarray} 
where we use the standard parameters $\sigma = 10$, $\beta = 8/3$ and $\rho = 28$. We simulate 
the dynamics by integrating these equations using the Runge-Kutta method with timesteps of 
$\Delta t = 0.02$. We use varying starting points on the attractor.\\
The equations are symmetric under the transformation $(x,y,z) \rightarrow (-x,-y,z)$.
Thus the mirror-attractor differs only in the z-coordinate. Furthermore, the mean of the attractor 
in z-direction is far away from the origin at $\bar{z} \approx 23.5$. This makes it an especially useful example for breaking the 
symmetry in the reservoir.\\
As a secondary test case we use the Halvorsen equations \cite{sprott2003chaos}

\begin{eqnarray}
\begin{aligned}
\dot x &= -\sigma x - 4y - 4z - y^2 \\
\dot y &= -\sigma y - 4z - 4x - z^2 \\
\dot z &=  -\sigma z - 4x - 4y - x^2 \ .
\label{eq:halvorsen}
\end{aligned}
\end{eqnarray} 
with $\sigma = 1.3$. We simulate the dynamics in the same way as for the Lorenz equations. This 
system also exhibits chaotic behavior but does not have any symmetries under inversion of its 
coordinates. It has a cyclic symmetry, which should however not be relevant in this context.
\\
Both Lorenz and Halvorsen system are 3-dimensional autonomous dissipative flows.

%%%%%%%%%%%%%%%%%%%%%%%%%%%%%%%

\subsection{Reservoir Computing and Simple ESN}
\label{sec:Reservoir Computing}

There is a multitude of ways to design an ESN. In this paper we use several different variants,
which will be introduced in the following sections. In general the input is fed into the reservoir
and influences its dynamics. A usually linear readout is trained to translate the state of the reservoir 
into the desired output. The reservoir state is then a random, high-dimensional, nonlinear transformation
of all previous input data. This naturally gives it a kind of memory.\\

In an ESN the dynamics of the reservoir are generally governed by an update equation for the reservoir 
state 
$\textbf{r}_t \in \mathbb{R}^{N}$ of the form
\begin{eqnarray}
\textbf{r}_{t+1} = f(\textbf{A}\textbf{r}_{t}, \textbf{W}_{in} \textbf{x}_{t}) 
\label{eq:general updating}
\end{eqnarray} 
Here $f$ is called activation function, the adjacency matrix 
$\textbf{A} \in \mathbb{R}^{N \times N} $ represents the network, $\textbf{x}_{t} \in \mathbb{R}^{d_{x}}$ 
is the input fed into the reservoir and $\textbf{W}_{in} \in \mathbb{R}^{d_{x} \times N}$ is the input
matrix. There are many possible choices for $f$ and ways to construct $\textbf{A}$ and $\textbf{W}_{in}$.
Here we always create $\textbf{A}$ as an Erd\"os-Renyi random network. Sparse 
networks have been found to be advantageous\cite{griffith2019forecasting}. The weights of the network are then drawn 
uniformly from $[-1,1]$ and afterwards rescaled to fix the spectral radius $\rho$ to some fixed value.
$\rho$ is a free hyperparameter.\\
We chose $\textbf{W}_{in}$ to be also sparse, in the sense that every row has only one nonzero element.
This means every reservoir node is only connected to one degree of freedom of the input \cite{lu2018attractor}.
We fixed the number of nodes per dimension to be the same plus or minus one. The 
nonzero elements are drawn uniformly from the interval $[-1,1]$ and then rescaled with a factor
$s_{input}$, which is 
another free hyperparameter. \\
From this we can then compute the output $\textbf{y}_{t} \in \mathbb{R}^{d_{y}}$. Here we are
interested in the prediction case, where we train the ESN to approximate $\textbf{y}_{t} \approx 
\textbf{x}_{t+1}$. Thus, the dimension of input and output are the same, so we use $d_{x} = d_{y} := d$. 
The readout is characterized by 

\begin{eqnarray}
\textbf{y}_{t} = \textbf{W}_{out}\tilde{\textbf{r}}_{t}
\label{eq:general readout}
\end{eqnarray} 

where typically $\tilde{\textbf{r}}_{t} = \textbf{r}_{t}$, but  $\tilde{\textbf{r}}_{t} \in 
\mathbb{R}^{\tilde{N}}$ can also be some nonlinear transformation or extension of $\textbf{r}_{t}$.
The readout matrix $\textbf{W}_{out} \in \mathbb{R}^{d \times \tilde{N}}$ is the only part of the
ESN that is trained. This is typically done via simple Ridge Regression \cite{hoerl1970ridge}. \\
To train the reservoir the 
training data $\textbf{x}^{train} = \{\textbf{x}_{0},... , \textbf{x}_{T_{train}}\}$ is 
fed into the reservoir to get the sequence $\textbf{r}^{train} = \{\textbf{r}_{0},... , \textbf{r}_{T_{train}+1}\}$.
The first $T_{sync}$ timesteps of $\textbf{r}$ are then discarded. This transient period is only used 
to synchronize the reservoir with the training data. This frees the ESN of any influence of the 
reservoir's initial condition thanks to the \textit{fading memory property}. The state of a 
properly designed reservoir continuously loses its dependence on past states over time. For a detailed
description see e.g. \cite{maass02, grigoryeva2018echo, boyd1985fading}
\\

% Fig %%%
\begin{figure}[!htbp] 
  \begin{center}
    \includegraphics[width=1.0\linewidth]{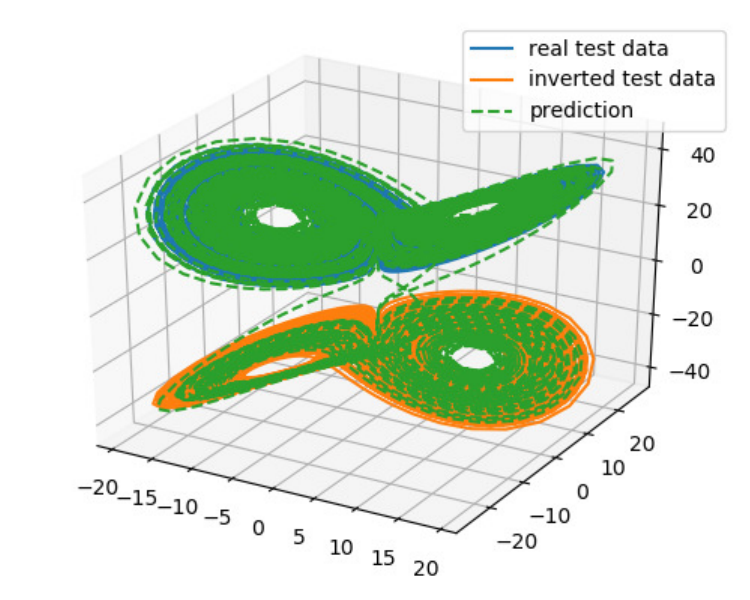}
    \caption{Failed prediction of the Lorenz system with a simple ESN. The trajectory jumps down to
    the mirror-attractor.}
    \label{fig:fail1}
    \includegraphics[width=1.0\linewidth]{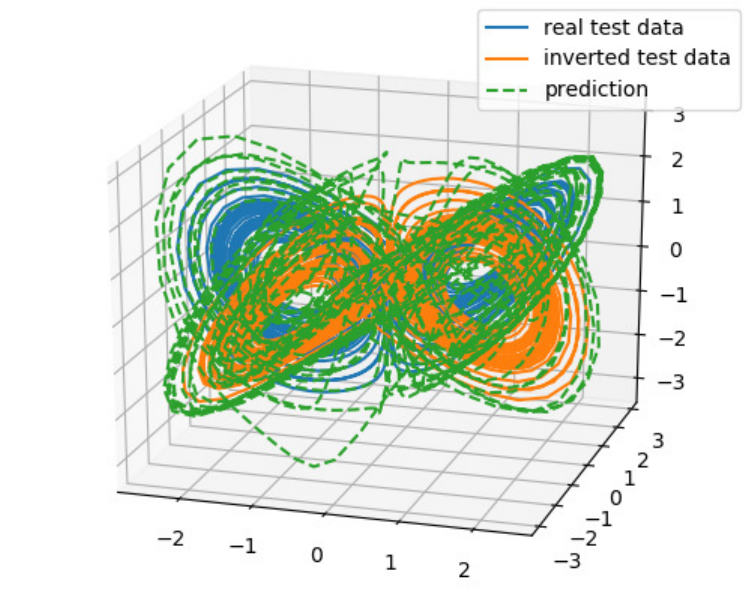}
    \caption{Failed prediction of zero-mean, normalized Lorenz data. The trajectory jumps frequently between
    the original and the  mirror-attractor.}
    \label{fig:fail2}
  \end{center}
\end{figure}

Now the readout matrix can be calculated by minimizing 
\begin{eqnarray}
\sum^{}_{T_{sync} \leq t \leq T_{train}} {\parallel  \textbf{W}_{out}\tilde{\textbf{r}}_{t} - 
\textbf{v}_{t} \parallel}^2 - \beta {\parallel \textbf{W}_{out} \parallel}^2 \ ,
\label{eq:minimizing}
\end{eqnarray} 
where we get another hyperparameter $\beta$ from the regularization. The target output $\textbf{v}_{t}$
is in the case of prediction just $\textbf{x}_{t+1}$. We thus get \cite{hoerl1970ridge}
\begin{eqnarray}
\textbf{W}_{out} = (\tilde{\textbf{r}}^{T} \tilde{\textbf{r}} + \beta \mathds{1})^{-1}
\tilde{\textbf{r}}^{T} \textbf{v} \ ,
\label{eq:ridgematrix}
\end{eqnarray} 
where $\textbf{r}$ is  $\textbf{r}^{train}$ in matrix form after discarding the synchronization 
steps and $\textbf{v}$ is analogous.\\
In the following we always use a network with $N=200$, $T_{train}=10500$ and $T_{sync}=500$ and 
average degree $k=4$ unless otherwise stated. The spectral radius $\rho$, the regularization 
parameter $\beta$ and the input scaling 
$s_{input}$ are optimized for specific problems.
\\

Our basic setup is close to what Jaeger \cite{jaeger2001echo} originally proposed and it is one of the most widely used 
variants. We call it simple ESN because all other designs we use are extensions of it. It is 
defined by the following equations:
\begin{eqnarray}
\textbf{r}_{t+1} = \tanh(\textbf{A}\textbf{r}_{t} +  \textbf{W}_{in} \textbf{x}_{t}) 
\label{eq:simple updating}
\end{eqnarray} 
\begin{eqnarray}
\textbf{y}_{t} = \textbf{W}_{out}\textbf{r}_{t} 
\label{eq:simple readout}
\end{eqnarray} 
The activation function is a sigmoidal function, specifically a hyperbolic tangent, which is the typical 
choice. The reservoir states are not transformed before the readout.
With this setup successful predictions of different datasets have been made in many cases. However, 
we can sometimes see very specific ways in which they fail as illustrated in figure \ref{fig:fail1} and 
figure \ref{fig:fail2}. When predicting the Lorenz attractor (see Sec. \ref{sec:Lorenz}), the prediction 
sometimes jumps to an inverted version of the training dataset, which we call \textit{mirror-attractor}.
To investigate the severity of the problem, we created 1000 realizations with predictions of  
500000 timesteps. We found that in 98.5\% of cases, where the prediction crossed the zero 
in the $z$-direction,
this lead to a jump to the other attractor. 18\% of the predictions did not exhibit a jump at any point. Among 
those that did, the average time of the first jump was about 31000 timesteps (539 Lyapunov times). Typically, the trajectory changed
from one attractor to the other multiple times.

\begin{figure}[!htbp] 
  \begin{center}
    \includegraphics[width=1.0\linewidth]{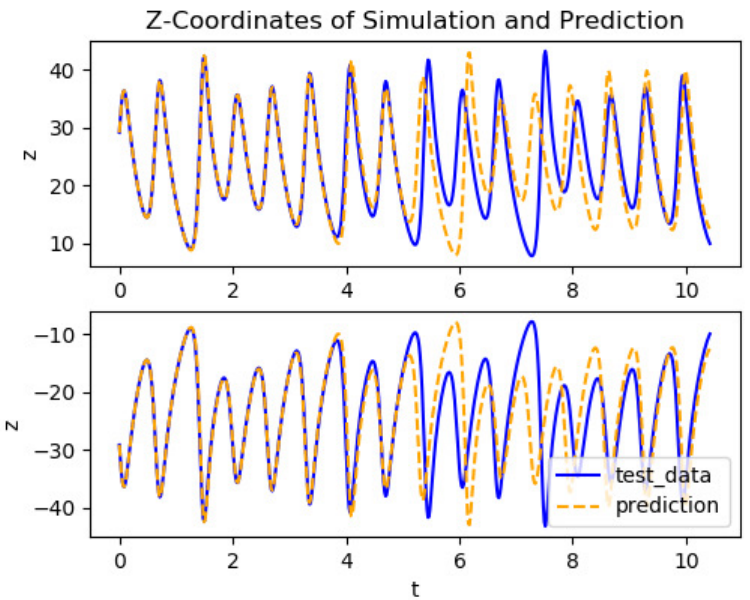}
    \caption{Prediction of the $z$-coordinate after synchronization with original data (upper) and inverted data 
    (lower) for the simple ESN. The $t$-axis is given in units of Lyapunov times.}
    \label{fig:siminvert}
  \end{center}
\end{figure}

While this is concerning, it still allows for decent short-time predictions. We were able to reach average forecast
horizons of about 400 timesteps (7 Lyapunov times) 
after hyperparameter optimization. However, when the data is brought to zero-mean the ability to make accurate 
predictions largely breaks down. After hyperparameter optimization we get a forecast horizon of
about 90 timesteps (1.6 Lyapunov times). Since the two attractors overlap, the prediction jumps between them very 
frequently as we can see in figure \ref{fig:fail2}. Sometimes it even travels outside of both for a short 
time.
Since this kind of preprocessing is considered good practice in machine learning and usually leads to 
better results, this shows a severe failure of the method.
\\
Difficulties in predicting the Lorenz system, which is a widespread test case for ESNs, when using 
this kind of setup have been noted by previous studies. They were linked to the symmetry of the Lorenz 
equation under the transformation $(x,y,z) \rightarrow (-x,-y,z)$. However, we observe 
these problems with other datasets as well. We can largely explain this phenomenon by
mathematical analysis independent of the input data. 
\\
To prove this we do the following: Assume $\textbf{r}_0 = \textbf{0}$ w.l.o.g. because of the fading 
memory property. Let us now analyze
what happens, when instead of the original training sequence  
$\textbf{x}^{train} = \{\textbf{x}_{0},... , \textbf{x}_{T_{train}}\}$ we use its inverted version
 $-\textbf{x}^{train} = \{-\textbf{x}_{0},... , -\textbf{x}_{T_{train}}\}$ to train the readout matrix:

\begin{eqnarray}
\textbf{r}_{0}(-\textbf{x}^{train}) = \textbf{0} = -\textbf{r}_{0}(\textbf{x}^{train})
\end{eqnarray}
\begin{eqnarray}
\textbf{r}_{1}(-\textbf{x}^{train}) & = & \tanh( - \textbf{W}_{in} \textbf{x}_{n}) \\
& = & -\tanh(\textbf{W}_{in} \textbf{x}_{n}) \\
& = & -\textbf{r}_{1}(\textbf{x}^{train})
\label{eq:base case}
\end{eqnarray} 
This serves as the base case for our mathematical induction. We follow up with the induction step.
Assume
\begin{eqnarray}
\textbf{r}_{t}(-\textbf{x}^{train}) = -\textbf{r}_{t}(\textbf{x}^{train})
\label{eq:reservoirsymmetry}
\end{eqnarray} 
Then
\begin{eqnarray}
\textbf{r}_{t+1}(-\textbf{x}^{train}) &=& \tanh(\textbf{A}\textbf{r}_{t}(-\textbf{x}^{train}) - 
\textbf{W}_{in} \textbf{x}_{t})\\
&=& -\tanh(\textbf{A}\textbf{r}_{t}(\textbf{x}^{train}) + 
\textbf{W}_{in} \textbf{x}_{t})  \\
&=& -\textbf{r}_{t+1}(\textbf{x}^{train}) 
\label{induction step}
\end{eqnarray} 

Overall we get $\textbf{r}^{train}(-\textbf{x}^{train})=-
\textbf{r}^{train}(\textbf{x}^{train})$.
So the dynamics of the reservoir only changed sign and are otherwise unaffected. This is a 
consequence of the antisymmetry of the 
hyperbolic tangent. \\
Obviously, because of the linearity of the readout, we also get
\begin{eqnarray}
\textbf{y}_{t}(-\textbf{x}^{train}) = -\textbf{W}_{out} 
\textbf{r}_{t}(\textbf{x}^{train})=-\textbf{y}_{t}(\textbf{x}^{train})
\end{eqnarray}
And finally
\begin{eqnarray}
\textbf{W}_{out}(-\textbf{x}) &=& 
(\textbf{r}^{T}(-\textbf{x}) \textbf{r}(-\textbf{x})+
\beta \mathds{1})^{-1}\textbf{r}^{T}(-\textbf{x})(-\textbf{x}) \\
&=& (\textbf{r}^{T}(\textbf{x}) \textbf{r}(\textbf{x})+
\beta \mathds{1})^{-1}\textbf{r}^{T}(\textbf{x})\textbf{x} \\
&=& \textbf{W}_{out}(\textbf{x})
\label{eq:ridgematrix symmetry}
\end{eqnarray} 
So training the simple ESN with inverted data is equivalent to training on the original data and both
lead to learning the exact same parameters. Thus, it can never map these sequences to either the same 
output or any output that differs by anything other than the sign. It is therefore not universal. It can 
only fully learn the dynamics of systems that are themselves point symmetric at the origin.\\
Furthermore, when we use a simple ESN for prediction it is now obvious that it learns to replicate the inverted
mirror-attractor as well as the real attractor. In cases where they overlap they are however incompatible.
When they do not overlap, but are close
enough to each other this makes jumps possible.
\\
In figure \ref{fig:siminvert} this is demonstrated by comparing the predictions
of an already trained simple ESN after being synchronized with additional Lorenz data either unchanged 
or inverted. We can see that the prediction of inverted data is simply the inversion of the prediction
of the original data, just as expected from theory.
\\
For a jump to happen, the reservoir has to arrive at a state that matches better with the 
mirror-attractor than with the real one. Since the reservoir has memory it is not obvious how fast
input data from the phase-space region of the mirror-attractor can actually make that happen. 
Empirically we found that crossing the zero in the z-direction leads to a jump in $98.5\%$ of 
cases. We further observed that inverting the input in a single 
timestep was reliably enough to push the prediction on the mirror-attractor.
This implies a strong sensitivity to the input data with regards to inducing jumps.
\\
It is clear that this kind of symmetry creates a significant limitation for
the simple ESN. We can therefore easily explain the previous problems with this kind of approach as well
as the so far mostly empirical success of some methods combating them. In many recent 
publications the readout was extended with a some kind of nonlinear transformation. The empirical 
advantage of this has been explored without theoretical explanation by Chattopadhyay et al. 
\cite{chattopadhyay2019data}. Typically quadratic terms are included in the readout (see Sec. 
\ref{sec:Lu Readout}). This was to our knowledge originally introduced
by Lu et al. \cite{lu2017reservoir} in
order to specifically solve a problem relating to the symmetry of the Lorenz equations, when using 
the ESN as an observer. It has since been used successfully in many more general cases without 
theoretical explanation. From our analysis it is now clear that this readout breaks the antisymmetry
of the ESN as a whole, which is completely independent of any symmetries of the input data.
\\
A similar analysis can be fruitful on many different designs of ESN. For example in a recent Paper 
by Carroll and Pecora \cite{carroll2019network} the following update equation was used:
\begin{eqnarray}
r_{i}(t+1) & = & \alpha r_{i}(t) + (1-\alpha) \tanh( \\
& & {} \sum_{j=1}^{M} A_{ij} r_j(t) + w_i x(t) + 1)
\label{eq:Carrol}
\end{eqnarray}

Where $w_i$ are the elements of what they call \textit{input vector}. Empirically they found that the 
performance suffered for $w_i = 1$  $\forall  i$ compared to $w_i \in \{+1,-1\}$. We can explain this
with a symmetry under the transformation $s(t) \rightarrow -s(t) - \frac{2}{w}$ for any constant 
$w_i = w$ $\forall i$ similar to what we found for the simple ESN. As soon as $w_i$ takes on different 
values for different $i$ this symmetry is broken.

\subsection{Breaking Symmetry}
\label{sec:Breaking}
To better understand what is the best way to break the harmful antisymmetry in the simple ESN, we 
test four different designs. There are two main ways of approaching the problem. We can either break
the symmetry in the reservoir, e.g by changing the activation function, or in the readout.
When choosing the latter option, equation \ref{eq:reservoirsymmetry} still holds. The dynamics of the
reservoir still do not change meaningfully with the sign of the training data. Only during prediction
does the influence of the readout actually come into play.
\\
We use two designs following each approach.

\subsubsection{Output Bias}
\label{sec:Output Bias}

\begin{figure}[!htbp] 
  \begin{center}
    \includegraphics[width=1.0\linewidth]{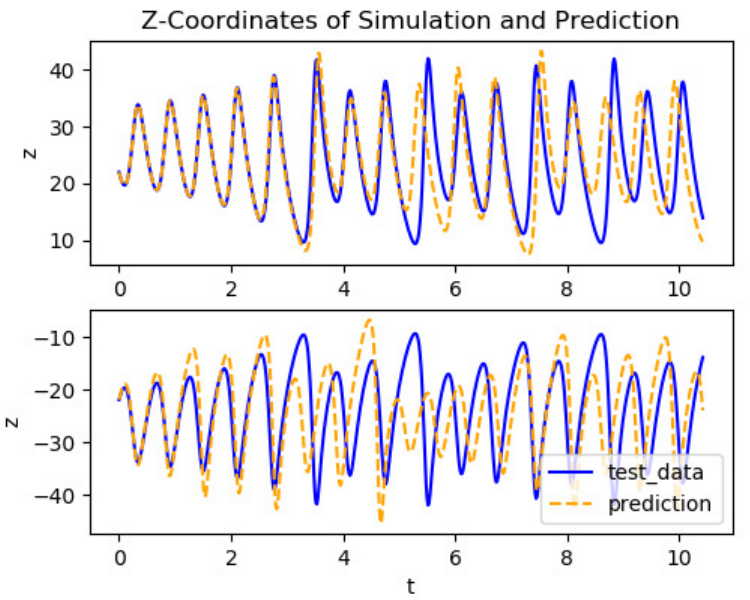}
    \caption{Prediction of the $z$-coordinate after synchronization with original data (upper) and 
    inverted data (lower) for the ESN with output bias. The $t$-axis is given in units of Lyapunov times. 
    The prediction of inverted data is
    perturbed but still in the mirror-attractor.}
    \label{fig:obinvert}
  \end{center}
\end{figure}

This is one of the simplest ways to break the symmetry. The readout is changed by using 
$\tilde{\textbf{r}} = \{r_1,r_2,...,r_N,1\}$. Effectively this leads to
\begin{eqnarray}
\textbf{y}_{t} = \textbf{W}_{out}\tilde{\textbf{r}}_{t} = 
\tilde{\textbf{W}}_{out}\textbf{r}_{t} + \textbf{b}
\label{eq:bias readout}
\end{eqnarray} 
where $\textbf{b} \in \mathbb{R}^{d}$ is called \textit{bias-term} and is fixed in the Linear 
Regression. This very basic extension of Linear Regression is often already seen as good practice.
Formally this breaks the symmetry.
\begin{eqnarray}
\textbf{y}_{t}(-\textbf{r}_t)=
-\tilde{\textbf{W}}_{out}\textbf{r}_{t} + \textbf{b}=
-\textbf{y}_{t}(\textbf{r}_t)+2\textbf{b}
\label{eq:bias symmetry}
\end{eqnarray} 
we note however that this way there are only $d$ parameters to represent the difference under 
sign-change.

\subsubsection{Lu Readout}
\label{sec:Lu Readout}

\begin{figure}[!htbp]
  \begin{center}
    \includegraphics[width=1\linewidth]{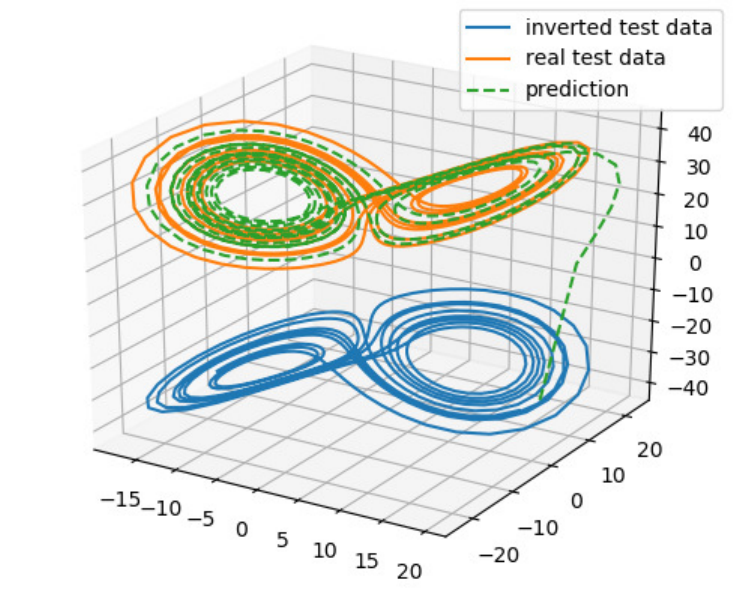}
    \caption{Prediction after synchronizing with inverted data for the ESN with Lu readout.}
    \label{fig:luinvert}
  \end{center}
\end{figure}

As previously mentioned, a quadratic extension of the readout has recently become popular 
after its introduction by Lu et al. \cite{lu2017reservoir} We use two different
variants of this approach. In its most powerful version it consists of using 
$\tilde{\textbf{r}} = \{r_1,r_2,...,r_N,r_1^2,r_2^2,...,r_N^2\}$ in the readout.\\
Effectively we get
\begin{eqnarray}
\textbf{y}_{t} = \textbf{W}_{out}\tilde{\textbf{r}}_{t} = 
\textbf{W}^1_{out}\textbf{r}_{t} + \textbf{W}^2_{out}\textbf{r}^2_{t} \ .
\label{eq:lu readout}
\end{eqnarray} 
Where $\textbf{W}_{out} \in \mathbb{R}^{d \times 2N}$ can be divided in 
$\textbf{W}^1_{out}$ and $\textbf{W}^2_{out} \in \mathbb{R}^{d \times N}$
And

\begin{eqnarray}
\textbf{y}_{t}(-\textbf{r}_t) &=&
-\textbf{W}^1_{out}\textbf{r}_{t} + \textbf{W}^2_{out}\textbf{r}^2_{t} \\
&=& -\textbf{y}_{t}(\textbf{r}_t)+2 \textbf{W}^2_{out}\textbf{r}^2_{t} \ .
\label{eq:lu symmetry}
\end{eqnarray} 
The number of parameters that represent the difference under sign-change is $d \times N$.
In the following we will call this \textit{extended Lu readout}. We use this in order to test
the full potential of this approach. However, the higher number of parameters in the output
matrix makes a quantitative comparison to other approaches unfair. For this purpose we also test 
a second version.
\\
In the original work by Lu et al. only half of the nodes are squared in the readout and each is 
only used either in its linear or in its quadratic form. This can be achieved with a transformation
of the reservoir state like
$\tilde{\textbf{r}} = \{r_1,r_2^2,r_3,r_4^2,...,r_{N-1},r_N^2\}$, where we assumed $N$ to be even.
We call this \textit{Lu readout} or \textit{regular Lu readout}. This way the number of parameters
is unaffected, which makes a fair comparison with the other methods possible.

\subsubsection{Input Shift}
\label{sec:Input Shift}

This design for an ESN has been proven to be universal by Grigoryeva and Ortega \cite{grigoryeva2018echo} Specifically
this means it can approximate any causal and time-invariant filter with the fading-memory property,
which naturally excludes any problems with symmetry. This is also recommended in ``A practical guide to
Applying Echo State Networks'' by Luko{\v{s}}evi{\v{c}}ius \cite{lukosevicius2012}.
\\
For this design the activation function of the simple ESN is extended by including a random bias term
in every node of the reservoir. It can be written as a random vector $\gamma \in \mathbb{R}^N$ and
gives the following new update equation:
\begin{eqnarray}
\textbf{r}_{t+1} = \tanh(\textbf{A}\textbf{r}_{t} +  \textbf{W}_{in} \textbf{x}_{t} + \gamma) \ .
\label{eq:Input Shift updating}
\end{eqnarray} 
The readout is unchanged from the simple ESN. Unlike the first two methods this breaks the symmetry in
the reservoir itself.\\
We draw the elements of $\gamma$ uniformly from $[-s_{\gamma};,s_{\gamma}]$ where $s_{\gamma}$ is a 
new hyperparameter to be optimized. This was a somewhat arbitrary choice for simplicity. In 
principle we could instead use a normal distribution, the distribution of the training data, etc.

\subsubsection{Mixed Activation Functions}
\label{sec:Mix}
Another way of breaking the symmetry directly in the reservoir is to replace some of the odd $\tanh$
activation functions with even functions. This was inspired by a different framing of the problem:
Every node in the network can be understood as a function of the concatenation of input datum and 
reservoir state $\tilde{\textbf{x}} = \{r_1,...,r_N,x_1,..,x_d\} $. The readout is then simply a linear
combination of these functions, where the weights are optimized to approximate the output. The nodes
differ by the random parameters introduced through the elements of $\textbf{W}_{in}$, $\textbf{A}$ and 
$\gamma$, if input shift is included. In general we want this set of functions to approximate a basis
in the corresponding function space to be as powerful as possible. In the case of the simple ESN these 
functions are all odd and 
any linear combination of them will still be odd. Mixing in even functions gives us access to the 
whole function space as any function can be divided in an even and an odd part.
\\
As even function we simply used $\tanh^2$. We assigned half of the nodes connected to each input 
dimension to be even nodes, where this activation functions is used.

\begin{table}[!htbp]

\begin{tabular}{ l | l | l | l | l}
\hline
\hline
ESN design & F.H. in $\Delta t$ ($\tau_{\lambda}$) &
$\lambda \pm \sigma$ & $\nu \pm \sigma$ \\
\hline
Simple ESN & 90.9(1.6) & $0.2 \pm 0.2$ & $1.6 \pm 0.5$ \\
\hline
Output Bias &  149.7(2.6) & $0.3 \pm 0.2$ & $2.0 \pm 0.5$  \\
\hline
Mixed Activations  &  538.2(9.4) & $0.87 \pm 0.03$ & $1.97 \pm 0.13$  \\
\hline
Input Shift  &  629.3(11.0) & $0.87 \pm 0.02$ & $1.978\pm 0.008$  \\
\hline
Lu Readout  & 558.4(9.7) &
$0.87 \pm 0.04$ & $1.96 \pm 0.17$ \\
\hline
Ext. Lu Readout  &  631.3(11.0) & $0.87 \pm 0.02$ & $1.978 \pm 0.008$  \\
\hline
Test Data  &  $\infty$ & $0.87 \pm 0.02$ & $1.978 \pm 0.008$  \\
\hline
\hline
\end{tabular}

\caption{Performance of the different ESN designs on zero-mean Lorenz data. Comparison to the original 
Lorenz data in last row. Forecast horizon (F.H.) given in units of timesteps and Lyapunov times
in brackets.}
\label{tab:zeromean}
\end{table}

\section{Results}
\label{sec:Results}

\begin{figure}[!htbp]
	\begin{center}
		\includegraphics[width=1\linewidth]{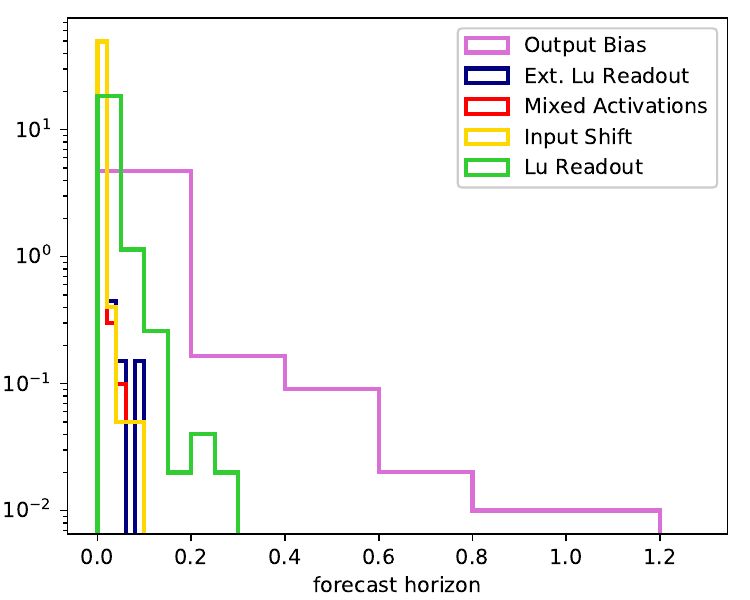}
		\caption{Normalized histogram of the forecast horizon (in units of Lyapunov times) with respect to the inverted test
		data after synchronizing with inverted training data for the four different 
		symmetry-breaking designs.  }
		\label{fig:invertdist}
	\end{center}
\end{figure}
%%%%%%

\subsection{Predicting The Mirror-Attractor}
\label{sec:Mirror Attractor}

To test the ability of the four methods to break the symmetry of the simple ESN, we tried to force
them to predict the mirror-attractor of the Lorenz equations after being trained with regular data.
If the symmetry is truly broken, we expect this prediction to fail completely, indicating that the
ESN did not learn anything about the mirror-attractor. \\
To accomplish this we trained our ESNs with regular Lorenz data and then synchronized it with the 
inverted next 500 timesteps of the simulation. We then measured the forecast horizon of the prediction
in regards to the (also inverted) test data. For comparison, we also looked at the prediction after 
synchronization with the same data without inversion. \\
In figure \ref{fig:obinvert} we see the behavior of the ESN with output bias.
It differs from before in that the prediction of the mirror-attractor is not simply the 
inversion of the regular prediction as for the simple ESN in figure 
\ref{fig:siminvert}. However, even though it is generally a worse prediction, it clearly follows
the inverted 
trajectory. The ESN has still learned a slightly perturbed version of the mirror-attractor.
\\
When using input shift, Lu readout or mixed activations, we never observed a prediction staying 
in the vicinity of the mirror-attractor. Most of them instead leave it immediately and quickly
converge to the real Lorenz attractor
as in figure \ref{fig:luinvert}. In some cases the trajectory finds some other fixed point instead, but it never 
stays in the mirror-attractor. Qualitatively we get the same behavior when using mixed 
activations or input shift instead.
\\
Furthermore, we made 1000 predictions of the mirror-attractor with all four designs while varying 
the network and the starting point of the training data. The distribution of forecast horizons is
shown in figure \ref{fig:invertdist}. The output bias clearly sticks out as the only method showing
the ability to predict the mirror-attractor. Some realizations reach forecast horizons up to 
one Lyapunov time while the other methods never go beyond 0.1 Lyapunov times corresponding to 
only O(1) timesteps. An exception is the regular Lu readout, where the prediction
tends to stay on the mirror-attractor slightly longer than for input shift, extended Lu readout and
mixed activation functions. However, all of these trajectories converge to the original attractor 
afterwards.
\\
We also tested the rate of jumps between the attractors for the ESN with output bias analogously to the simple ESN
in Sec. \ref{sec:Reservoir Computing} by making 1000 predictions with 500000 timesteps. Network and training
data were varied for each realization. This time 30.5\% of them did not show any jump. For the others the 
average time of the first jump was after about 33000 timesteps (574 Lyapunov times). 95.2\% of times the $z$-coordinate crossed zero
it lead to a jump.
This might indicate a small improvement, but the fundamental problem has not been solved by the 
output bias.\\
We did not observe any jumps when we used the other methods.

\subsection{Zero-mean Lorenz}
\label{sec:zeromean}
% Fig %%%
\begin{figure}[!htbp]
  \begin{center}
    \includegraphics[width=1\linewidth]{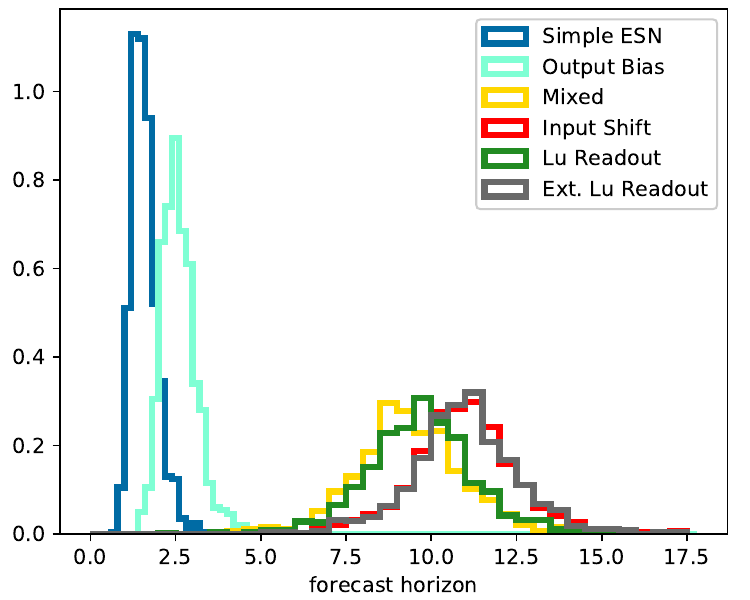}
    \caption{Normalized histogram of forecast horizons (in units of Lyapunov times) for normalized, zero-mean Lorenz data with all 
    five variants of ESN. Based on 1000 realizations (varying training data, network and 
    starting point of prediction) for each. }
    \label{fig:zeromeanfh}
  \end{center}
\end{figure}

%%%%%%
\begin{figure*}[!htbp] 
  \begin{center}
    \includegraphics[width=1.0\linewidth]{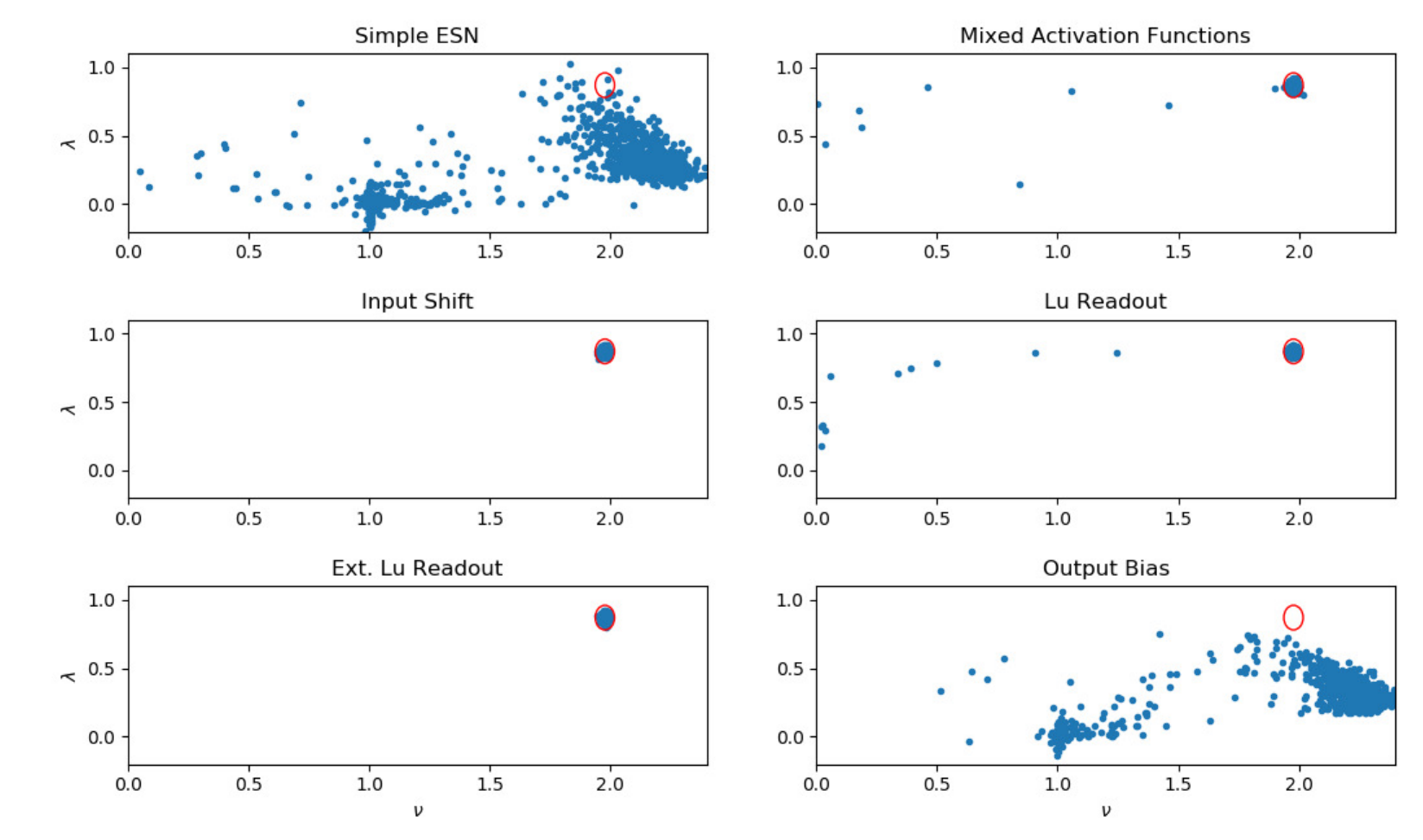}
    \caption{Scatterplot of the Largest Lyapunov Exponent against the correlation dimension
    when predicting zero-mean Lorenz data. Red ellipse corresponds to five times the standard deviation 
    of the test data. Based on 1000 realizations for each setup as in figure
    \ref{fig:zeromeanfh}.}
    \label{fig:scatter}
  \end{center}
\end{figure*}
%%%%%%
To further compare the performance of the different ESNs, we test the ability to learn and predict
Lorenz data where the mean has been shifted to the origin. As already discussed, this leads to 
overlap between real and mirror-attractor. So even though this preprocessing is usually preferred,
it makes the simple ESN's problems with antisymmetry especially severe. We also rescaled all 
data to have a standard deviation of 1.\\
To get reliable quantitative results, we first carried out a hyperparameter optimization on this task 
for every design used. We used a grid search with 100 realizations for each point in parameter space.
The best parameters are chosen on the basis of the highest average forecast horizon. Further details
and results can be found in the appendix.
\\
With the optimized hyperparameters we created 1000 realizations for each design and measured 
forecast horizon, largest Lyapunov exponent and correlation dimension. To accurately represent the climate 
we used predictions with a length of 20000 timesteps. The results are compiled in table 
\ref{tab:zeromean} and figure \ref{fig:zeromeanfh} and \ref{fig:scatter}. 
\\
The simple
ESN's forecast horizon consistently lies in a region below 3.5 Lyapunov times with a mean of 1.6
and similarly to the 
first task the output bias offers a noticeable but small
improvement with a mean of 2.6 Lyapunov times. The other four methods to break the symmetry all 
seem to work in principle. Their forecast horizons mostly lie between 
7 and 14 Lyapunov times. Extended Lu readout and 
input shift show no significant difference with an average forecast horizon of about 11 
Lyapunov times, 
while mixed activation functions and regular Lu readout show a lower average   
forecast horizon of 9.4 and 9.6 respectively. \\
In agreement with the results for the forecast horizon, the 
simple ESN's climate produces values of largest Lyapunov exponent and correlation dimension
far away from the desired region. Again the output bias is only a small improvement. All results
for extended Lu readout and input shift lie in the direct vicinity of the target and the mean values and 
standard deviations match those of the test data within uncertainty. Regular Lu readout and
mixed activation functions also reproduce the correct mean values but with much higher variance.
This can be attributed to the clear outliers visible in the plots. We note that 
the results for the climate are less reliable than those for the forecast horizon since we did
not optimize the hyperparameters for this task.
\\
In some cases the predicted trajectory diverged completely or got stuck in a fixed point. 
This made the proper calculation of the largest Lyapunov exponent impossible and thus these values are
excluded from that statistic. This happened 6 times with the simple ESN setup, 30 times with mixed
activation functions and 5 times, when using a regular Lu readout. Input shift, extended Lu readout
and even output bias did not show this behavior in any prediction in this experiment. 
\\
Overall using a regular Lu readout or 
mixed activation functions did not perform quite as well on this standard task as input shift and
extended Lu readout.

\subsection{Halvorsen and Lorenz}
\label{sec:halorenz}

\begin{figure}[!htbp]
  \begin{center}
    \includegraphics[width=1\linewidth]{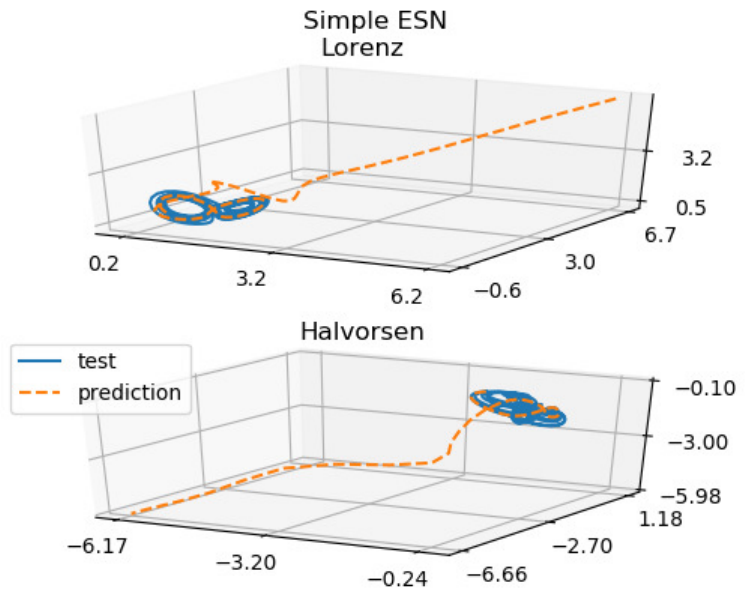}
    \caption{Example of predictions after training the same network on Lorenz and Halvorsen data 
    simultaneously as described in \ref{sec:halorenz}. Here we use the simple ESN setup.}
    \label{fig:simhalorenz}
  \end{center}
\end{figure}

\begin{figure}[!htbp]
  \begin{center}
    \includegraphics[width=1\linewidth]{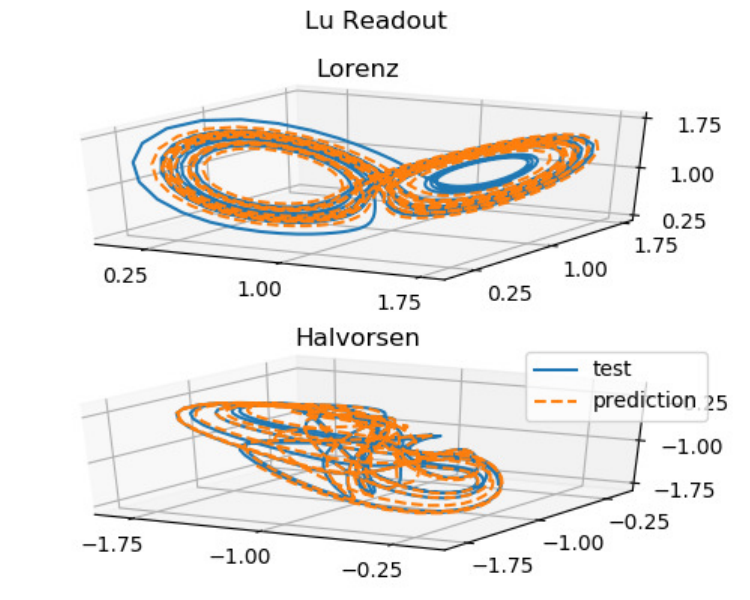}
    \caption{
    Example prediction of Lorenz and Halvorsen attractor with a single ESN with extended Lu readout.}
    \label{fig:luhalorenz}
  \end{center}
\end{figure}

Finally, we compare the five different ESNs on a task that we specifically designed to test their 
symmetry breaking abilities. For this goal we create a dataset by simulating both the Lorenz and the 
Halvorsen system. The mean of the Lorenz data is shifted to $(1,1,1)$ and the mean of the 
Halvorsen data is shifted to $(-1,-1,-1)$. Both are rescaled so that no datapoint has a distance
higher than 1 from the mean in any dimension. This ensures that the two attractors do not overlap
while lying completely in the region of each others mirror-attractor. To train the ESN to 
simultaneously be able to predict both systems we use the following trick. Firstly, we synchronize it
with the Lorenz data and record $\textbf{r}_{Lorenz}^{train}$ after the initial transient period.
We do however not calculate $W_{out}$ yet. Instead we repeat the process with the Halvorsen data. 
Now the transient period has the additional use of letting the reservoir forget about the Lorenz 
system. This way we get $\textbf{r}_{Lorenz}^{train}$ and  $\textbf{r}_{Halvorsen}^{train}$. We 
simply concatenate them to get a single dataset $\textbf{r}^{train}$, from which we finally compute 
the readout matrix. As desired output we use an analogous concatenation of the Lorenz data and the 
Halvorsen data. We note that, since the linear readout is in no way sensitive to the causal 
relationship between the reservoir states and the transient period at the second training stage 
was discarded, the transition between the two systems in itself does not influence training.\\ 
This way the ESN has to learn 
dynamics that are governed by a completely different set of equations instead of the mirror-attractor.
In the end it should be able to predict both attractors depending on the starting point of
the reservoir states. Since this is a more difficult task and to make sure that possible failures
are not just due to a lack of nodes we use $N=500$ in this experiment. However, we made similar 
qualitative observations for smaller and larger networks.
\\
To be able to do a quantitative analysis, we first performed a hyperparameter optimization 
as described in the appendix. We used the product of forecast horizons on both systems as a measure
of performance. Afterwards we carried out the same experiment as in Sec. \ref{sec:zeromean} with
this combined dataset. We always made a prediction on both attractors with the same network. The 
results are compiled in table \ref{tab:halorenz}.\\
Unsurprisingly we observe that the simple ESN is not able to master this task 
(see figure \ref{fig:simhalorenz}). Most predictions completely diverge from both attractors. In the 
handful of cases where one of the predictions actually reproduced the climate of one attractor, 
the other one was always a complete failure.
Qualitatively we see the same results when including an output bias.
\\
Since all predictions with the simple ESN and almost all with the ESN with output bias either 
diverged or converged to some fixed point, we could not provide meaningful results for the climate.
We note however that the short-term prediction of the 
Lorenz system was actually significantly better for both than in Sec. \ref{sec:zeromean}. We 
attribute this to the higher number of nodes. Still the inability to reproduce long-term behavior
indicates that this kind of problem can only be solved with a properly broken symmetry as we assumed.
\\
Again the other methods to break the symmetry are all successful (see as an example figure
\ref{fig:luhalorenz}) and predict both attractors with the same training quite well. The results in the forecast horizon for input shift, mixed 
activation functions and the regular Lu readout are very similar with an average of about 9.8
Lyapunov times (Lorenz) and 10.7 Lyapunov times (Halvorsen). However, only the input shift was
able to reproduce the climate
with the same accuracy as the test data. 
\\
It is notable that the extended Lu readout performed significantly better than the others
in terms of the forecast horizon on both systems. The averages were 10.6 Lyapunov times (Lorenz)
and 11.6 Lyapunov times (Halvorsen). In contrast, there was also a small number
 of predictions that completely diverged or got stuck in a fixed point with this design.
This occurred 26 times with the extended Lu readout and 8 times with the regular Lu readout. 
As in Sec. \ref{sec:zeromean} these predictions are not included in the Lyapunov exponent
statistic. The same did not happen
when
using input shift or mixed activation functions. This might be due to the fact that the Lu readout
does not break the symmetry in the reservoir itself. For this more complicated task the additional
parameters in the readout might not always be sufficient to encode the difference in the dynamics
for 
a sign change. It could be related to the fact that those dynamics are completely independent of
each other.

\begin{table}[!htbp]

\begin{tabular}{ l | l | l | l | l}
\hline
\hline
ESN design & F.H. in $\Delta t$ ($\tau_{\lambda}$) &
$\lambda \pm \sigma$ & $\nu \pm \sigma$ \\
\hline
Simple ESN & 183.2(3.2) & - & - \\
{}         & 54.2(0.8)  & - & - \\
\hline
Output Bias & 226.3(3.9)  & - & - \\
{}         & 57.7(0.9)  & - & - \\
\hline
Mixed Activations  &  563.4(9.8) & $0.87 \pm 0.03$ & $1.99 \pm 0.02$  \\
{} 		   &  709.4(10.5) & $0.74 \pm 0.03$ & $1.88 \pm 0.04$  \\ 
\hline
Input Shift  &  570.5(9.9) & $0.87 \pm 0.02$ & $1.992\pm 0.007$  \\
{}	     &  721.4(10.7) & $0.74 \pm 0.03$ & $1.88 \pm 0.04$   \\
\hline
Lu Readout  &  561.8(9.7) & $0.87 \pm 0.02$ & $1.97 \pm 0.18$  \\
{}          &  719.4(10.7) & $0.74 \pm 0.03$ & $1.87 \pm 0.04$ \\
\hline
Ext. Lu Readout  &  610.9(10.6) & $0.87 \pm 0.05$ & $1.93 \pm 0.35$  \\
{}          &  784.1(11.6) & $0.74 \pm 0.03$ & $1.87 \pm 0.03$ \\

\hline
Test Data  &  $\infty$ & $0.87 \pm 0.02$ & $1.993 \pm 0.007$  \\
{}         &  $\infty$ & $0.74 \pm 0.03$ & $1.87 \pm 0.04$    \\
\hline
\hline
\end{tabular}

\caption{Performance of the different ESN designs on combined Lorenz and Halvorsen data. Upper value is always 
Lorenz and lower Halvorsen. Comparison to the original data in last row. Forecast horizon (F.H.) given in units of timesteps and Lyapunov times
in brackets.}
\label{tab:halorenz}
\end{table}

%%%%%%

\section{Discussion}
\label{sec:Conclusion}

In the present work we showed a mathematical proof for the antisymmetry of the simple ESN with 
regards to changing the sign of the input. This is a consequence of the antisymmetry of the 
activation function. It makes it impossible to fully learn the dynamics 
of any attractor that is not point symmetric around the origin. In practice we observed that the 
prediction jumps to an inverted version of the real attractor we call mirror-attractor. This is 
especially disastrous if the two overlap. From this we conclude that this setup is not suitable 
for general tasks and should not be used.\\
Furthermore, we note that the sensitivity to this kind of symmetries with regards to the input 
is a universal property of ESNs and Reservoir Computers in general. This is in no way limited to 
the specifics of the simple ESN. It must be kept in mind in every reservoir design and 
can explain the empirical success or failure of some of them.\\
In our experiments with the output bias we found that formally breaking the symmetry alone is not
enough to solve the problems associated with it. It was only able to improve the performance marginally
and we still observed the appearance of an only slightly perturbed mirror-attractor. This might be 
due to the fact that the number of parameters representing the symmetry break in this approach 
is too low to accurately model the difference.\\
We were however able to successfully break the symmetry and solve the problem with three other approaches: Introducing an input shift in the activation function, using a 
mixture of even and odd activation functions and including squared nodes in the readout.
All of them were able to eliminate the mirror-attractor and make qualitatively good predictions even
for zero-mean Lorenz data, where the overlap with the mirror-attractor is a severe problem for the 
simple ESN. They were further all able to master the task of predicting a dataset made of Lorenz and Halvorsen
data, where it was necessary to learn completely different dynamics in the regime of the 
mirror-attractor.
\\
The input shift proved in our test as a very useful and reliable tool to break
the symmetry. The performance in all tasks was consistently good in short-time prediction.
It successfully reproduced the climate statistics of the test data for the zero-mean Lorenz dataset
and even for the combined Lorenz and Halvorsen data.
This method was the only one to never produce outliers of completely failed
predictions. It is also worth stressing that to our knowledge universality is only proven for 
ESNs with input shift. One disadvantage of this approach is the additional hyperparameter, which 
has to be optimized.
\\
Even though the Lu readout in its regular form also seems to have broken the symmetry successfully,
the results were generally not quite on par with the input shift.  However, this gap was
bridged by the extended Lu readout, which poses a convenient way to add more parameters to the model
without increasing the size of the reservoir. For the zero-mean Lorenz data this lead to a performance
essentially identical with that of the input shift. In the case of the combined dataset the short-time
prediction surpassed the input shift, while the climate was
not reproduced with the same accuracy. This
improvement is likely due to the higher number of parameters and thus higher complexity 
in the readout. While this is of course accompanied by an increase in computational time, it 
demonstrates the general possibility to enhance the prediction abilities of a given reservoir by 
extending the readout. Even though the regular Lu readout seems to be enough to break the symmetry, 
using even higher order nonlinear transformations of nodes and an even bigger output matrix 
could further increase the performance. An application of this could be found in physical RC, 
where the dynamics
of the reservoir might be inaccesible.
One might however consider making the dynamics of the reservoir more complex, while keeping 
the simple linear readout, to be more in line with the philosophy of RC.
\\
The ESN with mixed activation functions performed similarly well to the ESN with regular Lu readout.
This is interesting, because the former can be understood as using squared nodes not only in the
readout,
but also in the dynamics of the reservoir. The results imply that this does not lead to a meaningful
improvement. At least for our implementation we also found it to
have a higher time cost. Thus, we do not recommend its use in the given form. However, the usage of
different functions, different ratios, etc. could lead to better performance. Further research in this 
direction is needed.
\\
In light of these results we recommend both the input shift and the Lu readout as methods 
to break the symmetry.
%%%%%%%%%%%%%%%%%%%%%%%%%%%%%%%%%%%%%%%%%%%%%%%%%%%%%%%%%%%%%%%%%%%%%%%%%%%%%%%%

\section*{Acknowledgements}
We wish to acknowledge useful discussions and comments from Jonas Aumeier, Sebastian Baur, 
Youssef Mabrouk, Alexander Haluszczynski and Hubertus Thomas. We also want to thank our anonymous
reviewers for their helpful suggestions.

\section*{Data Availability}
The data that support the findings of this study are available from the corresponding author
upon reasonable request.

\appendix
\section{Hyperparameter Optimization}

The Hyperparameter Optimization was carried out as a simple grid search with the aim to maximize the 
forecast horizon. For the simple ESN we searched over $a$ and $\epsilon$, with 
\begin{eqnarray}
s_{input} &=& a(1-\epsilon) \\ 
\rho &=& a\epsilon 
\label{eq:parameterisation}
\end{eqnarray}.
The same was done for the ESN with output bias and the ESN with Lu readout. For the ESN with input 
shift we additionally varied the scale $s_\gamma$ and for the mixed activations we replaced $a$ with $a_1$
and $a_2$, which were optimized for the $\tanh$-nodes and the $\tanh^2$-nodes separately.

\subsection{Predicting the mirror-attractor}

Since we were less interested in quantitative results in this case we did not perform a real hyperparameter
optimization procedure. Instead we used the parameters from our previous work \cite{haluszczynski2020reducing}
for the simple ESN and manually searched the parameters for the others to reproduce the Lorenz
attractor reasonably well. This left us with the parameters in table \ref{tab:mirror}.

\begin{table}[H]
\setlength{\medmuskip}{0mu}

\begin{tabular}{ l | l | l | l | l | l | l}
\hline
\hline
{} & Simple &
Out. Bias & Lu readout & Mixed & Inp. Shift \\
\hline
$a$ & 0.32 & 0.32  & 0.32 & - & 0.32 \\
\hline
$\epsilon$ &  0.5 & 0.5 & 0.5 & 0.5 & 0.5 \\
\hline
$\beta$ & $1.9\times10^{-11}$ & $1.9\times10^{-11}$ & $1.9\times 10^{-11}$ & $1.9\times10^{-11}$ & $1.9\times10^{-11}$ \\
\hline
$s_{\gamma}$ &  - & - & - & -& 13.  \\

\hline
$a_1$ &  - & - & - & 0.32 & - \\
\hline
$a_2$ &  - & - & - & 0.32 & -  \\

\hline
\hline
\end{tabular}

\caption{Hyperparameter choices for the Prediction of the mirror-attractor.}
\label{tab:mirror}
\end{table}

\subsection{Zero-mean Lorenz} 
In the case
of the zero-mean Lorenz data we simulated 
100 trajectories of training data and 100 trajectories of test data. At every point in hyperparameter space
we generate a new network and a new $\textbf{W}_{in}$ for each trajectory. We then choose the 
hyperparameters with the highest average forecast horizon.\\
Since it did not seem to depend strongly on the other hyperparameters, the problem or the specific design, 
we simply set $\beta=1.9 \times 10^{-11}$ as in the first task to save time with the already very costly grid search.
\\
The results are compiled in tables \ref{tab:OptZeroSimple}, \ref{tab:OptZeroOB}, \ref{tab:OptZeroLU}, \ref{tab:OptZeroELu}, 
\ref{tab:OptZeroInput} and \ref{tab:OptZeroMix}.

\begin{table}[H]

\begin{tabular}{ l | l | l | l | l | l}
\hline
\hline
Hyperparameter & Min &
Max & Step Size & Optimal \\
\hline
$a$ & 0.1 & 3.0  & 0.1 & 1.0 \\
\hline
$\epsilon$ &  0.0 & 1.0 & 0.05 & 0.7  \\

\hline
\hline
\end{tabular}

\caption{Hyperparameter range and results for the simple ESN on zero-mean Lorenz data.}
\label{tab:OptZeroSimple}
\end{table}

\begin{table}[H]

\begin{tabular}{ l | l | l | l | l | l}
\hline
\hline
Hyperparameter & Min &
Max & Step Size & Optimal \\
\hline
$a$ & 0.1 & 3.0  & 0.1 & 1.0 \\
\hline
$\epsilon$ &  0.0 & 1.0 & 0.05 & 0.7  \\

\hline
\hline
\end{tabular}

\caption{Hyperparameter range and results for the ESN with output bias on zero-mean Lorenz data.}
\label{tab:OptZeroOB}
\end{table}

\begin{table}[H]

\begin{tabular}{ l | l | l | l | l | l}
\hline
\hline
Hyperparameter & Min &
Max & Step Size & Optimal \\
\hline
$a$ & 0.1 & 3.0  & 0.1 & 0.9 \\
\hline
$\epsilon$ &  0.0 & 1.0 & 0.05 & 0.55  \\

\hline
\hline
\end{tabular}

\caption{Hyperparameter range and results for the ESN with regular Lu readout on zero-mean Lorenz data.}
\label{tab:OptZeroLU}
\end{table}

\begin{table}[H]

\begin{tabular}{ l | l | l | l | l | l}
\hline
\hline
Hyperparameter & Min &
Max & Step Size & Optimal \\
\hline
$a$ & 0.1 & 3.0  & 0.1 & 1.3 \\
\hline
$\epsilon$ &  0.0 & 1.0 & 0.05 & 0.4  \\

\hline
\hline
\end{tabular}

\caption{Hyperparameter range and results for the ESN with extended Lu readout on zero-mean Lorenz data.}
\label{tab:OptZeroELu}
\end{table}

\begin{table}[H]

\begin{tabular}{ l | l | l | l | l | l}
\hline
\hline
Hyperparameter & Min &
Max & Step Size & Optimal \\
\hline
$a$ & 0.2 & 3.0  & 0.2 & 1.2 \\
\hline
$\epsilon$ &  0.0 & 1.0 & 0.1 & 0.6  \\
\hline
$s_{\gamma}$  &  0.3 &  3.0 & 0.3 & 1.5  \\

\hline
\hline
\end{tabular}

\caption{Hyperparameter range and results for the ESN with input shift on zero-mean Lorenz data.}
\label{tab:OptZeroInput}
\end{table}

\begin{table}[H]

\begin{tabular}{ l | l | l | l | l | l}
\hline
\hline
Hyperparameter & Min &
Max & Step Size & Optimal \\
\hline
$a_1$  &  0.2 & 3.0 & 0.2 & 0.6  \\
\hline
$a_2$  &  0.2 & 3.0 & 0.2 & 0.8  \\
\hline
$\epsilon$ &  0.1 & 1.0 & 0.1 & 0.6  \\
\hline

\hline
\hline
\end{tabular}

\caption{Hyperparameter range and results for the ESN with mixed activation functions on 
zero-mean Lorenz data.}
\label{tab:OptZeroMix}
\end{table}

\subsection{Halvorsen and Lorenz} 

For the combined dataset of Halvorsen and Lorenz attractor we simulate 100 training and test 
trajectories of each system and use them as described in Sec. \ref{sec:halorenz}. As before we train 
a completely new reservoir on each trajectory for every point in parameter space. For every 
realization the product of the two forecast horizons is calculated. The optimal hyperparameters are 
chosen to maximize this this product averaged over the trajectories.\\
For this task we did include the regularization parameter $\beta$ in the search with a logarithmic
scale from $10^{-13}$ to 0.001 in 11 steps. We consistently found $\beta = 10^{-10}$ to be the best
choice for all designs.  
\\
The results are compiled in tables \ref{tab:OptHLSimple}, \ref{tab:OptHLOB}, \ref{tab:OptHLLu}, \ref{tab:OptHLELu}, 
\ref{tab:OptHLInput} and \ref{tab:OptHLMix}.

\begin{table}[H]

\begin{tabular}{ l | l | l | l | l | l}
\hline
\hline
Hyperparameter & Min &
Max & Step Size & Optimal \\
\hline
$a$ & 1.1 & 4.0  & 0.1 & 2.0 \\
\hline
$\epsilon$ &  0.0 & 1.0 & 0.05 & 0.4  \\

\hline
\hline
\end{tabular}

\caption{Hyperparameter range and results for the simple ESN on combined Lorenz and Halvorsen data.}
\label{tab:OptHLSimple}
\end{table}

\begin{table}[H]

\begin{tabular}{ l | l | l | l | l | l}
\hline
\hline
Hyperparameter & Min &
Max & Step Size & Optimal \\
\hline
$a$ & 1.1 & 4.0  & 0.1 & 1.9 \\
\hline
$\epsilon$ &  0.0 & 1.0 & 0.05 & 0.4  \\

\hline
\hline
\end{tabular}

\caption{Hyperparameter range and results for the ESN with output bias on combined Lorenz and Halvorsen data.}
\label{tab:OptHLOB}
\end{table}

\begin{table}[H]

\begin{tabular}{ l | l | l | l | l | l}
\hline
\hline
Hyperparameter & Min &
Max & Step Size & Optimal \\
\hline
$a$ & 1.1 & 4.0  & 0.1 & 1.9 \\
\hline
$\epsilon$ &  0.0 & 1.0 & 0.05 & 0.3  \\

\hline
\hline
\end{tabular}

\caption{Hyperparameter range and results for the ESN with regular Lu readout on combined Lorenz and Halvorsen data.}
\label{tab:OptHLLu}
\end{table}

\begin{table}[H]

\begin{tabular}{ l | l | l | l | l | l}
\hline
\hline
Hyperparameter & Min &
Max & Step Size & Optimal \\
\hline
$a$ & 1.1 & 4.0  & 0.1 & 2.3 \\
\hline
$\epsilon$ &  0.0 & 1.0 & 0.05 & 0.45  \\

\hline
\hline
\end{tabular}

\caption{Hyperparameter range and results for the ESN with extended Lu readout on combined Lorenz and Halvorsen data.}
\label{tab:OptHLELu}
\end{table}

\begin{table}[H]

\begin{tabular}{ l | l | l | l | l | l}
\hline
\hline
Hyperparameter & Min &
Max & Step Size & Optimal \\
\hline
$a$ & 1.2 & 4.0  & 0.2 & 3.0 \\
\hline
$\epsilon$ &  0.0 & 1.0 & 0.1 & 0.1  \\
\hline
$s_{\gamma}$  &  0.3 &  3.0 & 0.3 & 1.5  \\

\hline
\hline
\end{tabular}

\caption{Hyperparameter range and results for the ESN with input shift on combined Lorenz and Halvorsen data.}
\label{tab:OptHLInput}
\end{table}

\begin{table}[H]

\begin{tabular}{ l | l | l | l | l | l}
\hline
\hline
Hyperparameter & Min &
Max & Step Size & Optimal \\
\hline
$a_1$  &  1.2 & 4.0 & 0.2 & 2.8  \\
\hline
$a_2$  &  1.2 & 4.0 & 0.2 & 2.4  \\
\hline
$\epsilon$ &  0.1 & 1.0 & 0.1 & 0.2  \\
\hline

\hline
\hline
\end{tabular}

\caption{Hyperparameter range and results for the ESN with mixed activation functions on 
combined Lorenz and Halvorsen data.}
\label{tab:OptHLMix}
\end{table}

\section*{References}

\providecommand{\noopsort}[1]{}\providecommand{\singleletter}[1]{#1}%


\begin{thebibliography}{26}%
\makeatletter
\providecommand \@ifxundefined [1]{%
 \@ifx{#1\undefined}
}%
\providecommand \@ifnum [1]{%
 \ifnum #1\expandafter \@firstoftwo
 \else \expandafter \@secondoftwo
 \fi
}%
\providecommand \@ifx [1]{%
 \ifx #1\expandafter \@firstoftwo
 \else \expandafter \@secondoftwo
 \fi
}%
\providecommand \natexlab [1]{#1}%
\providecommand \enquote  [1]{``#1''}%
\providecommand \bibnamefont  [1]{#1}%
\providecommand \bibfnamefont [1]{#1}%
\providecommand \citenamefont [1]{#1}%
\providecommand \href@noop [0]{\@secondoftwo}%
\providecommand \href [0]{\begingroup \@sanitize@url \@href}%
\providecommand \@href[1]{\@@startlink{#1}\@@href}%
\providecommand \@@href[1]{\endgroup#1\@@endlink}%
\providecommand \@sanitize@url [0]{\catcode `\\12\catcode `\$12\catcode
  `\&12\catcode `\#12\catcode `\^12\catcode `\_12\catcode `\%12\relax}%
\providecommand \@@startlink[1]{}%
\providecommand \@@endlink[0]{}%
\providecommand \url  [0]{\begingroup\@sanitize@url \@url }%
\providecommand \@url [1]{\endgroup\@href {#1}{\urlprefix }}%
\providecommand \urlprefix  [0]{URL }%
\providecommand \Eprint [0]{\href }%
\providecommand \doibase [0]{http://dx.doi.org/}%
\providecommand \selectlanguage [0]{\@gobble}%
\providecommand \bibinfo  [0]{\@secondoftwo}%
\providecommand \bibfield  [0]{\@secondoftwo}%
\providecommand \translation [1]{[#1]}%
\providecommand \BibitemOpen [0]{}%
\providecommand \bibitemStop [0]{}%
\providecommand \bibitemNoStop [0]{.\EOS\space}%
\providecommand \EOS [0]{\spacefactor3000\relax}%
\providecommand \BibitemShut  [1]{\csname bibitem#1\endcsname}%
\let\auto@bib@innerbib\@empty
%</preamble>
\bibitem [{\citenamefont {Tang}\ \emph {et~al.}(2020)\citenamefont {Tang},
  \citenamefont {Kurths}, \citenamefont {Lin}, \citenamefont {Ott},\ and\
  \citenamefont {Kocarev}}]{tang2020}%
  \BibitemOpen
  \bibfield  {author} {\bibinfo {author} {\bibfnamefont {Y.}~\bibnamefont
  {Tang}}, \bibinfo {author} {\bibfnamefont {J.}~\bibnamefont {Kurths}},
  \bibinfo {author} {\bibfnamefont {W.}~\bibnamefont {Lin}}, \bibinfo {author}
  {\bibfnamefont {E.}~\bibnamefont {Ott}}, \ and\ \bibinfo {author}
  {\bibfnamefont {L.}~\bibnamefont {Kocarev}},\ }\bibfield  {title} {\enquote
  {\bibinfo {title} {Introduction to focus issue: When machine learning meets
  complex systems: Networks, chaos, and nonlinear dynamics},}\ }\href@noop {}
  {\bibfield  {journal} {\bibinfo  {journal} {Chaos: An Interdisciplinary
  Journal of Nonlinear Science}\ }\textbf {\bibinfo {volume} {30}},\ \bibinfo
  {pages} {063151} (\bibinfo {year} {2020})}\BibitemShut {NoStop}%
\bibitem [{\citenamefont {Lu}, \citenamefont {Hunt},\ and\ \citenamefont
  {Ott}(2018)}]{lu2018attractor}%
  \BibitemOpen
  \bibfield  {author} {\bibinfo {author} {\bibfnamefont {Z.}~\bibnamefont
  {Lu}}, \bibinfo {author} {\bibfnamefont {B.~R.}\ \bibnamefont {Hunt}}, \ and\
  \bibinfo {author} {\bibfnamefont {E.}~\bibnamefont {Ott}},\ }\bibfield
  {title} {\enquote {\bibinfo {title} {Attractor reconstruction by machine
  learning},}\ }\href@noop {} {\bibfield  {journal} {\bibinfo  {journal}
  {Chaos: An Interdisciplinary Journal of Nonlinear Science}\ }\textbf
  {\bibinfo {volume} {28}},\ \bibinfo {pages} {061104} (\bibinfo {year}
  {2018})}\BibitemShut {NoStop}%
\bibitem [{\citenamefont {Pathak}\ \emph {et~al.}(2017)\citenamefont {Pathak},
  \citenamefont {Lu}, \citenamefont {Hunt}, \citenamefont {Girvan},\ and\
  \citenamefont {Ott}}]{pathak2017using}%
  \BibitemOpen
  \bibfield  {author} {\bibinfo {author} {\bibfnamefont {J.}~\bibnamefont
  {Pathak}}, \bibinfo {author} {\bibfnamefont {Z.}~\bibnamefont {Lu}}, \bibinfo
  {author} {\bibfnamefont {B.~R.}\ \bibnamefont {Hunt}}, \bibinfo {author}
  {\bibfnamefont {M.}~\bibnamefont {Girvan}}, \ and\ \bibinfo {author}
  {\bibfnamefont {E.}~\bibnamefont {Ott}},\ }\bibfield  {title} {\enquote
  {\bibinfo {title} {Using machine learning to replicate chaotic attractors and
  calculate lyapunov exponents from data},}\ }\href@noop {} {\bibfield
  {journal} {\bibinfo  {journal} {Chaos: An Interdisciplinary Journal of
  Nonlinear Science}\ }\textbf {\bibinfo {volume} {27}},\ \bibinfo {pages}
  {121102} (\bibinfo {year} {2017})}\BibitemShut {NoStop}%
\bibitem [{\citenamefont {Vlachas}\ \emph {et~al.}(2019)\citenamefont
  {Vlachas}, \citenamefont {Pathak}, \citenamefont {Hunt}, \citenamefont
  {Sapsis}, \citenamefont {Girvan}, \citenamefont {Ott},\ and\ \citenamefont
  {Koumoutsakos}}]{vlachas2019backpropagation}%
  \BibitemOpen
  \bibfield  {author} {\bibinfo {author} {\bibfnamefont {P.~R.}\ \bibnamefont
  {Vlachas}}, \bibinfo {author} {\bibfnamefont {J.}~\bibnamefont {Pathak}},
  \bibinfo {author} {\bibfnamefont {B.~R.}\ \bibnamefont {Hunt}}, \bibinfo
  {author} {\bibfnamefont {T.~P.}\ \bibnamefont {Sapsis}}, \bibinfo {author}
  {\bibfnamefont {M.}~\bibnamefont {Girvan}}, \bibinfo {author} {\bibfnamefont
  {E.}~\bibnamefont {Ott}}, \ and\ \bibinfo {author} {\bibfnamefont
  {P.}~\bibnamefont {Koumoutsakos}},\ }\href@noop {} {\enquote {\bibinfo
  {title} {Backpropagation algorithms and reservoir computing in recurrent
  neural networks for the forecasting of complex spatiotemporal dynamics},}\ }
  (\bibinfo {year} {2019}),\ \Eprint {http://arxiv.org/abs/1910.05266}
  {arXiv:1910.05266 [eess.SP]} \BibitemShut {NoStop}%
\bibitem [{\citenamefont {Chattopadhyay}\ \emph {et~al.}(2019)\citenamefont
  {Chattopadhyay}, \citenamefont {Hassanzadeh}, \citenamefont {Subramanian},\
  and\ \citenamefont {Palem}}]{chattopadhyay2019data}%
  \BibitemOpen
  \bibfield  {author} {\bibinfo {author} {\bibfnamefont {A.}~\bibnamefont
  {Chattopadhyay}}, \bibinfo {author} {\bibfnamefont {P.}~\bibnamefont
  {Hassanzadeh}}, \bibinfo {author} {\bibfnamefont {D.}~\bibnamefont
  {Subramanian}}, \ and\ \bibinfo {author} {\bibfnamefont {K.}~\bibnamefont
  {Palem}},\ }\bibfield  {title} {\enquote {\bibinfo {title} {Data-driven
  prediction of a multi-scale lorenz 96 chaotic system using a hierarchy of
  deep learning methods: Reservoir computing, ann, and rnn-lstm},}\ }\href@noop
  {} {\  (\bibinfo {year} {2019})}\BibitemShut {NoStop}%
\bibitem [{\citenamefont {Maass}, \citenamefont {Natschlaeger},\ and\
  \citenamefont {Markram}(2002)}]{maass02}%
  \BibitemOpen
  \bibfield  {author} {\bibinfo {author} {\bibfnamefont {W.}~\bibnamefont
  {Maass}}, \bibinfo {author} {\bibfnamefont {T.}~\bibnamefont {Natschlaeger}},
  \ and\ \bibinfo {author} {\bibfnamefont {H.}~\bibnamefont {Markram}},\
  }\bibfield  {title} {\enquote {\bibinfo {title} {Real-time computing without
  stable states: A new framework for neural computation based on
  perturbations},}\ }\href {\doibase 10.1162/089976602760407955} {\bibfield
  {journal} {\bibinfo  {journal} {Neural Computation}\ }\textbf {\bibinfo
  {volume} {14}},\ \bibinfo {pages} {2531--2560} (\bibinfo {year}
  {2002})}\BibitemShut {NoStop}%
\bibitem [{\citenamefont {Jaeger}(2001)}]{jaeger2001echo}%
  \BibitemOpen
  \bibfield  {author} {\bibinfo {author} {\bibfnamefont {H.}~\bibnamefont
  {Jaeger}},\ }\bibfield  {title} {\enquote {\bibinfo {title} {The “echo
  state” approach to analysing and training recurrent neural networks-with an
  erratum note},}\ }\href@noop {} {\bibfield  {journal} {\bibinfo  {journal}
  {Bonn, Germany: German National Research Center for Information Technology
  GMD Technical Report}\ }\textbf {\bibinfo {volume} {148}},\ \bibinfo {pages}
  {13} (\bibinfo {year} {2001})}\BibitemShut {NoStop}%
\bibitem [{\citenamefont {Van~der Sande}, \citenamefont {Brunner},\ and\
  \citenamefont {Soriano}(2017)}]{van2017advances}%
  \BibitemOpen
  \bibfield  {author} {\bibinfo {author} {\bibfnamefont {G.}~\bibnamefont
  {Van~der Sande}}, \bibinfo {author} {\bibfnamefont {D.}~\bibnamefont
  {Brunner}}, \ and\ \bibinfo {author} {\bibfnamefont {M.~C.}\ \bibnamefont
  {Soriano}},\ }\bibfield  {title} {\enquote {\bibinfo {title} {Advances in
  photonic reservoir computing},}\ }\href@noop {} {\bibfield  {journal}
  {\bibinfo  {journal} {Nanophotonics}\ }\textbf {\bibinfo {volume} {6}},\
  \bibinfo {pages} {561--576} (\bibinfo {year} {2017})}\BibitemShut {NoStop}%
\bibitem [{\citenamefont {Prychynenko}\ \emph {et~al.}(2018)\citenamefont
  {Prychynenko}, \citenamefont {Sitte}, \citenamefont {Litzius}, \citenamefont
  {Kr{\"u}ger}, \citenamefont {Bourianoff}, \citenamefont {Kl{\"a}ui},
  \citenamefont {Sinova},\ and\ \citenamefont
  {Everschor-Sitte}}]{prychynenko2018magnetic}%
  \BibitemOpen
  \bibfield  {author} {\bibinfo {author} {\bibfnamefont {D.}~\bibnamefont
  {Prychynenko}}, \bibinfo {author} {\bibfnamefont {M.}~\bibnamefont {Sitte}},
  \bibinfo {author} {\bibfnamefont {K.}~\bibnamefont {Litzius}}, \bibinfo
  {author} {\bibfnamefont {B.}~\bibnamefont {Kr{\"u}ger}}, \bibinfo {author}
  {\bibfnamefont {G.}~\bibnamefont {Bourianoff}}, \bibinfo {author}
  {\bibfnamefont {M.}~\bibnamefont {Kl{\"a}ui}}, \bibinfo {author}
  {\bibfnamefont {J.}~\bibnamefont {Sinova}}, \ and\ \bibinfo {author}
  {\bibfnamefont {K.}~\bibnamefont {Everschor-Sitte}},\ }\bibfield  {title}
  {\enquote {\bibinfo {title} {Magnetic skyrmion as a nonlinear resistive
  element: A potential building block for reservoir computing},}\ }\href@noop
  {} {\bibfield  {journal} {\bibinfo  {journal} {Physical Review Applied}\
  }\textbf {\bibinfo {volume} {9}},\ \bibinfo {pages} {014034} (\bibinfo {year}
  {2018})}\BibitemShut {NoStop}%
\bibitem [{\citenamefont {Tanaka}\ \emph {et~al.}(2019)\citenamefont {Tanaka},
  \citenamefont {Yamane}, \citenamefont {H{\'e}roux}, \citenamefont {Nakane},
  \citenamefont {Kanazawa}, \citenamefont {Takeda}, \citenamefont {Numata},
  \citenamefont {Nakano},\ and\ \citenamefont {Hirose}}]{tanaka2019recent}%
  \BibitemOpen
  \bibfield  {author} {\bibinfo {author} {\bibfnamefont {G.}~\bibnamefont
  {Tanaka}}, \bibinfo {author} {\bibfnamefont {T.}~\bibnamefont {Yamane}},
  \bibinfo {author} {\bibfnamefont {J.~B.}\ \bibnamefont {H{\'e}roux}},
  \bibinfo {author} {\bibfnamefont {R.}~\bibnamefont {Nakane}}, \bibinfo
  {author} {\bibfnamefont {N.}~\bibnamefont {Kanazawa}}, \bibinfo {author}
  {\bibfnamefont {S.}~\bibnamefont {Takeda}}, \bibinfo {author} {\bibfnamefont
  {H.}~\bibnamefont {Numata}}, \bibinfo {author} {\bibfnamefont
  {D.}~\bibnamefont {Nakano}}, \ and\ \bibinfo {author} {\bibfnamefont
  {A.}~\bibnamefont {Hirose}},\ }\bibfield  {title} {\enquote {\bibinfo {title}
  {Recent advances in physical reservoir computing: A review},}\ }\href@noop {}
  {\bibfield  {journal} {\bibinfo  {journal} {Neural Networks}\ }\textbf
  {\bibinfo {volume} {115}},\ \bibinfo {pages} {100--123} (\bibinfo {year}
  {2019})}\BibitemShut {NoStop}%
\bibitem [{\citenamefont {Griffith}, \citenamefont {Pomerance},\ and\
  \citenamefont {Gauthier}(2019)}]{griffith2019forecasting}%
  \BibitemOpen
  \bibfield  {author} {\bibinfo {author} {\bibfnamefont {A.}~\bibnamefont
  {Griffith}}, \bibinfo {author} {\bibfnamefont {A.}~\bibnamefont {Pomerance}},
  \ and\ \bibinfo {author} {\bibfnamefont {D.~J.}\ \bibnamefont {Gauthier}},\
  }\bibfield  {title} {\enquote {\bibinfo {title} {Forecasting chaotic systems
  with very low connectivity reservoir computers},}\ }\href@noop {} {\bibfield
  {journal} {\bibinfo  {journal} {Chaos: An Interdisciplinary Journal of
  Nonlinear Science}\ }\textbf {\bibinfo {volume} {29}},\ \bibinfo {pages}
  {123108} (\bibinfo {year} {2019})}\BibitemShut {NoStop}%
\bibitem [{\citenamefont {Carroll}\ and\ \citenamefont
  {Pecora}(2019)}]{carroll2019network}%
  \BibitemOpen
  \bibfield  {author} {\bibinfo {author} {\bibfnamefont {T.~L.}\ \bibnamefont
  {Carroll}}\ and\ \bibinfo {author} {\bibfnamefont {L.~M.}\ \bibnamefont
  {Pecora}},\ }\bibfield  {title} {\enquote {\bibinfo {title} {Network
  structure effects in reservoir computers},}\ }\href@noop {} {\bibfield
  {journal} {\bibinfo  {journal} {arXiv preprint arXiv:1903.12487}\ } (\bibinfo
  {year} {2019})}\BibitemShut {NoStop}%
\bibitem [{\citenamefont {Haluszczynski}\ and\ \citenamefont
  {R{\"a}th}(2019)}]{haluszczynski2019good}%
  \BibitemOpen
  \bibfield  {author} {\bibinfo {author} {\bibfnamefont {A.}~\bibnamefont
  {Haluszczynski}}\ and\ \bibinfo {author} {\bibfnamefont {C.}~\bibnamefont
  {R{\"a}th}},\ }\bibfield  {title} {\enquote {\bibinfo {title} {Good and bad
  predictions: Assessing and improving the replication of chaotic attractors by
  means of reservoir computing},}\ }\href@noop {} {\bibfield  {journal}
  {\bibinfo  {journal} {Chaos: An Interdisciplinary Journal of Nonlinear
  Science}\ }\textbf {\bibinfo {volume} {29}},\ \bibinfo {pages} {103143}
  (\bibinfo {year} {2019})}\BibitemShut {NoStop}%
\bibitem [{\citenamefont {Haluszczynski}\ \emph {et~al.}(2020)\citenamefont
  {Haluszczynski}, \citenamefont {Aumeier}, \citenamefont {Herteux},\ and\
  \citenamefont {R{\"a}th}}]{haluszczynski2020reducing}%
  \BibitemOpen
  \bibfield  {author} {\bibinfo {author} {\bibfnamefont {A.}~\bibnamefont
  {Haluszczynski}}, \bibinfo {author} {\bibfnamefont {J.}~\bibnamefont
  {Aumeier}}, \bibinfo {author} {\bibfnamefont {J.}~\bibnamefont {Herteux}}, \
  and\ \bibinfo {author} {\bibfnamefont {C.}~\bibnamefont {R{\"a}th}},\
  }\bibfield  {title} {\enquote {\bibinfo {title} {Reducing network size and
  improving prediction stability of reservoir computing},}\ }\href@noop {}
  {\bibfield  {journal} {\bibinfo  {journal} {Chaos: An Interdisciplinary
  Journal of Nonlinear Science}\ }\textbf {\bibinfo {volume} {30}},\ \bibinfo
  {pages} {063136} (\bibinfo {year} {2020})}\BibitemShut {NoStop}%
\bibitem [{\citenamefont {Carroll}(2020)}]{carroll2020path}%
  \BibitemOpen
  \bibfield  {author} {\bibinfo {author} {\bibfnamefont {T.}~\bibnamefont
  {Carroll}},\ }\bibfield  {title} {\enquote {\bibinfo {title} {Path length
  statistics in reservoir computers},}\ }\href@noop {} {\bibfield  {journal}
  {\bibinfo  {journal} {Chaos: An Interdisciplinary Journal of Nonlinear
  Science}\ }\textbf {\bibinfo {volume} {30}},\ \bibinfo {pages} {083130}
  (\bibinfo {year} {2020})}\BibitemShut {NoStop}%
\bibitem [{\citenamefont {Grassberger}\ and\ \citenamefont
  {Procaccia}(1983)}]{grassberger1983measuring}%
  \BibitemOpen
  \bibfield  {author} {\bibinfo {author} {\bibfnamefont {P.}~\bibnamefont
  {Grassberger}}\ and\ \bibinfo {author} {\bibfnamefont {I.}~\bibnamefont
  {Procaccia}},\ }\bibfield  {title} {\enquote {\bibinfo {title} {Measuring the
  strangeness of strange attractors},}\ }\href@noop {} {\bibfield  {journal}
  {\bibinfo  {journal} {Physica D: Nonlinear Phenomena}\ }\textbf {\bibinfo
  {volume} {9}},\ \bibinfo {pages} {189--208} (\bibinfo {year}
  {1983})}\BibitemShut {NoStop}%
\bibitem [{\citenamefont {{Grassberger}}(1983)}]{grassberger83a}%
  \BibitemOpen
  \bibfield  {author} {\bibinfo {author} {\bibfnamefont {P.}~\bibnamefont
  {{Grassberger}}},\ }\bibfield  {title} {\enquote {\bibinfo {title}
  {{Generalized dimensions of strange attractors}},}\ }\href {\doibase
  10.1016/0375-9601(83)90753-3} {\bibfield  {journal} {\bibinfo  {journal}
  {Physics Letters A}\ }\textbf {\bibinfo {volume} {97}},\ \bibinfo {pages}
  {227--230} (\bibinfo {year} {1983})}\BibitemShut {NoStop}%
\bibitem [{\citenamefont {Rosenstein}, \citenamefont {Collins},\ and\
  \citenamefont {De~Luca}(1993)}]{rosenstein1993practical}%
  \BibitemOpen
  \bibfield  {author} {\bibinfo {author} {\bibfnamefont {M.~T.}\ \bibnamefont
  {Rosenstein}}, \bibinfo {author} {\bibfnamefont {J.~J.}\ \bibnamefont
  {Collins}}, \ and\ \bibinfo {author} {\bibfnamefont {C.~J.}\ \bibnamefont
  {De~Luca}},\ }\bibfield  {title} {\enquote {\bibinfo {title} {A practical
  method for calculating largest lyapunov exponents from small data sets},}\
  }\href@noop {} {\bibfield  {journal} {\bibinfo  {journal} {Physica D:
  Nonlinear Phenomena}\ }\textbf {\bibinfo {volume} {65}},\ \bibinfo {pages}
  {117--134} (\bibinfo {year} {1993})}\BibitemShut {NoStop}%
\bibitem [{\citenamefont {Sandri}(1996)}]{sandri1996numerical}%
  \BibitemOpen
  \bibfield  {author} {\bibinfo {author} {\bibfnamefont {M.}~\bibnamefont
  {Sandri}},\ }\bibfield  {title} {\enquote {\bibinfo {title} {Numerical
  calculation of lyapunov exponents},}\ }\href@noop {} {\bibfield  {journal}
  {\bibinfo  {journal} {Mathematica Journal}\ }\textbf {\bibinfo {volume}
  {6}},\ \bibinfo {pages} {78--84} (\bibinfo {year} {1996})}\BibitemShut
  {NoStop}%
\bibitem [{\citenamefont {Lorenz}(1963)}]{lorenz1963deterministic}%
  \BibitemOpen
  \bibfield  {author} {\bibinfo {author} {\bibfnamefont {E.~N.}\ \bibnamefont
  {Lorenz}},\ }\bibfield  {title} {\enquote {\bibinfo {title} {Deterministic
  nonperiodic flow},}\ }\href@noop {} {\bibfield  {journal} {\bibinfo
  {journal} {Journal of the atmospheric sciences}\ }\textbf {\bibinfo {volume}
  {20}},\ \bibinfo {pages} {130--141} (\bibinfo {year} {1963})}\BibitemShut
  {NoStop}%
\bibitem [{\citenamefont {Sprott}\ and\ \citenamefont
  {Sprott}(2003)}]{sprott2003chaos}%
  \BibitemOpen
  \bibfield  {author} {\bibinfo {author} {\bibfnamefont {J.~C.}\ \bibnamefont
  {Sprott}}\ and\ \bibinfo {author} {\bibfnamefont {J.~C.}\ \bibnamefont
  {Sprott}},\ }\href@noop {} {\emph {\bibinfo {title} {Chaos and time-series
  analysis}}},\ Vol.~\bibinfo {volume} {69}\ (\bibinfo  {publisher}
  {Citeseer},\ \bibinfo {year} {2003})\BibitemShut {NoStop}%
\bibitem [{\citenamefont {Hoerl}\ and\ \citenamefont
  {Kennard}(1970)}]{hoerl1970ridge}%
  \BibitemOpen
  \bibfield  {author} {\bibinfo {author} {\bibfnamefont {A.~E.}\ \bibnamefont
  {Hoerl}}\ and\ \bibinfo {author} {\bibfnamefont {R.~W.}\ \bibnamefont
  {Kennard}},\ }\bibfield  {title} {\enquote {\bibinfo {title} {Ridge
  regression: Biased estimation for nonorthogonal problems},}\ }\href@noop {}
  {\bibfield  {journal} {\bibinfo  {journal} {Technometrics}\ }\textbf
  {\bibinfo {volume} {12}},\ \bibinfo {pages} {55--67} (\bibinfo {year}
  {1970})}\BibitemShut {NoStop}%
\bibitem [{\citenamefont {Grigoryeva}\ and\ \citenamefont
  {Ortega}(2018)}]{grigoryeva2018echo}%
  \BibitemOpen
  \bibfield  {author} {\bibinfo {author} {\bibfnamefont {L.}~\bibnamefont
  {Grigoryeva}}\ and\ \bibinfo {author} {\bibfnamefont {J.-P.}\ \bibnamefont
  {Ortega}},\ }\bibfield  {title} {\enquote {\bibinfo {title} {Echo state
  networks are universal},}\ }\href@noop {} {\bibfield  {journal} {\bibinfo
  {journal} {Neural Networks}\ }\textbf {\bibinfo {volume} {108}},\ \bibinfo
  {pages} {495--508} (\bibinfo {year} {2018})}\BibitemShut {NoStop}%
\bibitem [{\citenamefont {Boyd}\ and\ \citenamefont
  {Chua}(1985)}]{boyd1985fading}%
  \BibitemOpen
  \bibfield  {author} {\bibinfo {author} {\bibfnamefont {S.}~\bibnamefont
  {Boyd}}\ and\ \bibinfo {author} {\bibfnamefont {L.}~\bibnamefont {Chua}},\
  }\bibfield  {title} {\enquote {\bibinfo {title} {Fading memory and the
  problem of approximating nonlinear operators with volterra series},}\
  }\href@noop {} {\bibfield  {journal} {\bibinfo  {journal} {IEEE Transactions
  on circuits and systems}\ }\textbf {\bibinfo {volume} {32}},\ \bibinfo
  {pages} {1150--1161} (\bibinfo {year} {1985})}\BibitemShut {NoStop}%
\bibitem [{\citenamefont {Lu}\ \emph {et~al.}(2017)\citenamefont {Lu},
  \citenamefont {Pathak}, \citenamefont {Hunt}, \citenamefont {Girvan},
  \citenamefont {Brockett},\ and\ \citenamefont {Ott}}]{lu2017reservoir}%
  \BibitemOpen
  \bibfield  {author} {\bibinfo {author} {\bibfnamefont {Z.}~\bibnamefont
  {Lu}}, \bibinfo {author} {\bibfnamefont {J.}~\bibnamefont {Pathak}}, \bibinfo
  {author} {\bibfnamefont {B.}~\bibnamefont {Hunt}}, \bibinfo {author}
  {\bibfnamefont {M.}~\bibnamefont {Girvan}}, \bibinfo {author} {\bibfnamefont
  {R.}~\bibnamefont {Brockett}}, \ and\ \bibinfo {author} {\bibfnamefont
  {E.}~\bibnamefont {Ott}},\ }\bibfield  {title} {\enquote {\bibinfo {title}
  {Reservoir observers: Model-free inference of unmeasured variables in chaotic
  systems},}\ }\href@noop {} {\bibfield  {journal} {\bibinfo  {journal} {Chaos:
  An Interdisciplinary Journal of Nonlinear Science}\ }\textbf {\bibinfo
  {volume} {27}},\ \bibinfo {pages} {041102} (\bibinfo {year}
  {2017})}\BibitemShut {NoStop}%
\bibitem [{\citenamefont {Luko{\v{s}}evi{\v{c}}ius}(2012)}]{lukosevicius2012}%
  \BibitemOpen
  \bibfield  {author} {\bibinfo {author} {\bibfnamefont {M.}~\bibnamefont
  {Luko{\v{s}}evi{\v{c}}ius}},\ }\enquote {\bibinfo {title} {A practical guide
  to applying echo state networks},}\ in\ \href {\doibase
  10.1007/978-3-642-35289-8_36} {\emph {\bibinfo {booktitle} {Neural Networks:
  Tricks of the Trade: Second Edition}}},\ \bibinfo {editor} {edited by\
  \bibinfo {editor} {\bibfnamefont {G.}~\bibnamefont {Montavon}}, \bibinfo
  {editor} {\bibfnamefont {G.~B.}\ \bibnamefont {Orr}}, \ and\ \bibinfo
  {editor} {\bibfnamefont {K.-R.}\ \bibnamefont {M{\"u}ller}}}\ (\bibinfo
  {publisher} {Springer Berlin Heidelberg},\ \bibinfo {address} {Berlin,
  Heidelberg},\ \bibinfo {year} {2012})\ pp.\ \bibinfo {pages}
  {659--686}\BibitemShut {NoStop}%
\end{thebibliography}
\end{document}